 \documentclass[pdflatex,sn-mathphys]{sn-jnl}

\usepackage{lineno,hyperref}
\usepackage{algorithm}
\usepackage{algpseudocode}
\usepackage{caption}
\usepackage{subcaption}
\usepackage{multirow}
\usepackage{multicol}
\newcommand{\add}{\textcolor{gray}}

\usepackage{graphicx}
\graphicspath{ {./Figures/} }
\usepackage{lscape}

\jyear{2022}%

\theoremstyle{thmstyleone}%
\theoremstyle{thmstyletwo}%

\theoremstyle{thmstylethree}%

\begin{document}

\title[Usman et. al.]{Low-Latency Online Multiplier with Reduced Activities and Minimized Interconnect for Inner Product Arrays}


\author[1]{\fnm{Muhammad} \sur{Usman}}\email{usman@chosun.ac.kr}
\author[2]{\fnm{Milo\v{s}} \sur{D. Ercegovac}}\email{milos@cs.ucla.edu}

\author*[1]{\fnm{Jeong-A} \sur{Lee}}\email{jalee@chosun.ac.kr}

\affil[1]{\orgdiv{Department of Computer Engineering}, \orgname{Chosun University}, \orgaddress{\city{Gwangju}, \country{Republic of Korea}}}

\affil[2]{\orgdiv{Computer Science Department}, \orgname{University of California}, \orgaddress{\city{Los Angeles, CA}, \country{USA}}}


\abstract{Multiplication is indispensable and is one of the core operations in many modern applications including signal processing and neural networks. Conventional right-to-left (RL) multiplier extensively contributes to the power consumption, area utilization and critical path delay in such applications. This paper proposes a low latency multiplier based on online or left-to-right (LR) arithmetic which can increase throughput and reduce latency by digit-level pipelining. Online arithmetic enables overlapping successive operations regardless of data dependency because of the most significant digit first mode of operation. To produce most significant digit first, it uses redundant number system and we can have a carry-free addition, therefore, the delay of the arithmetic operation is independent of operand bit width.
The operations are  performed digit by digit serially from left to right which allows gradual increase in the slice activities making it suitable for implementation on reconfigurable devices. 
Serial nature of the online algorithm and gradual increment/decrement of active slices minimize the interconnects and signal activities resulting in overall reduction of area and power consumption. We present online multipliers with; both inputs in serial, and one in serial and one in parallel. Pipelined and non-pipelined designs of the proposed multipliers have been synthesized with GSCL 45nm technology on Synopsys Design Compiler. Thorough comparative analysis has been performed using widely used performance metrics. The results show that the proposed online  multipliers outperform the RL multipliers.}




\maketitle

\section{Introduction}\label{sec:introduction}
Multiplication is regarded as the fundamental operation in various signal processing and machine learning applications. Multipliers are regarded as the bottleneck in performance of various algorithms and it has been well versed that the architecture of the multiplier has direct effect on the performance of these applications in terms of area utilization, power consumption, and critical path delay. Based on the partial product generation and reduction, conventional multipliers can be categorized as linear array multipliers and tree multipliers \cite{liu2017design, lin2019novel}. Multipliers utilizing the digit-parallel computation require full bandwidth interconnection data-paths, resulting in an increased power/energy requirement. Digit-serial arithmetic is, therefore, often used to reduce the interconnection and hardware complexity. Both serial-parallel and serial-serial multipliers have been proposed over the years where one or both operands are provided serially, respectively. However, one of the drawbacks of using these multipliers is that their latency is dependent on the word size of operands. Furthermore, conventional arithmetic-based serial-serial and serial-parallel multipliers have limitation on throughput as the succeeding operation can be started only after the completion of current operation.

Online or left-to-right (LR) arithmetic \cite{ercegovac2017left}, which processes the input operands and generates result digits serially, most significant digit first (MSDF), can serve as a potential computing paradigm to address the limitations of conventional arithmetic and achieve aforementioned requirements of latency, throughput, area, and power. There are several benefits of using online arithmetic-based operators: 
\begin{itemize}
    \item Operands and result are streamed serially which reduces the interconnection bandwidth, area, and energy dissipation \cite{ercegovac2020}.
    \item The computation can be started without waiting for full precision data after an initial delay also called the online delay ($\delta$) during which, a few input bits are received. $\delta$ is a small number and represents the inter-operation latency.
    \item Successive operations have delay of $\delta+1$, therefore, they can be pipelined regardless of data dependency and achieve high throughput with minimum interconnect.
    \item It makes use of redundant number system which makes cycle time of the operation independent of precision.
    \item The working precision can be truncated such that $n$-digit precision result can be obtained by implementing $p<n$ digit slices.
    \item Computation can be stopped upon reaching the desired precision due to the MSDF mode of operation, thus, it can be configured as a variable/approximate computation algorithm. 
\end{itemize}

Online arithmetic has been widely adopted for the design and development of various simple and complex arithmetic circuits e.g., adders \cite{villalba2011radix}, multipliers \cite{elshafei2009hardware}, 3-D vector normalization \cite{huang2001fpga}, as well as in a number of compute intensive digital signal processing (DSP) applications \cite{galli2001design}, and matrix computation \cite{sinky2004design}. In \cite{zhao2016efficient}, online arithmetic operators were studied for implementation on FPGA where the property of online arithmetic was exploited and a fixed piece of hardware was utilized to perform calculations at any precision. This resulted $8\times$ speed-up to execute Newton's method compared to parallel-in-serial-out fixed point method. In \cite{shi2014efficient}, online operators were focused for efficient implementation on FPGA in order to achieve area savings and obtain speed-ups. For different online operators a reduction of upto $56\%$ silicon area and speed-up of around $1.5\times$ was shown on Xilinx Virtex-$6$ FPGA.

The idea of pipelining the online serial-serial multiplier had been presented in our previous work \cite{usman2021}.  In the continuation, we present the design of the online serial-serial pipelined multiplier with the derivations and detail of implementation of each module of the multiplier. Furthermore, the design of a pipelined serial-parallel online multiplier is also presented in this work. For the serial-serial online multiplier, we exploit the property of online arithmetic to truncate the maximum working precision i.e., to obtain $n$ bit result, $p<n$ bit slices are implemented which reduce the signal activities and interconnections. Furthermore, the slice activity of the proposed multiplier follows an increasing/decreasing pattern in terms of active slices i.e., the number of active slices increases systematically up-to a maximum of $p<n$ and then decreases. To which end, each step of the algorithm has been unrolled such that each stage instantiates only the desired number of bit slices to minimize the interconnection and signal activities.

The proposed designs of digit-level-pipelined serial-serial and serial-parallel multipliers have been synthesized to obtain area, power, and critical path delay results using Synopsys Design Compiler with GSCL $45$ nm liberty cell library from the Free45PDK and compared with non-pipelined versions of online multipliers as well as conventional multipliers.

The rest of the paper, we proceed as follows: an overview of online arithmetic and online multiplier has been presented in Section \ref{sec: online}. Details about the architecture, algorithm, and implementation of the proposed pipelined multipliers have been presented in Section \ref{sec: proposed}. The implementation results have been presented in Section \ref{sec: Results}, followed by the conclusion of the paper in Section \ref{sec: conclusion}.

\section{Online Arithmetic} \label{sec: online}

\begin{figure}[!ht]
	\begin{center}
\includegraphics[viewport=5 10 218 155,scale=1.0]{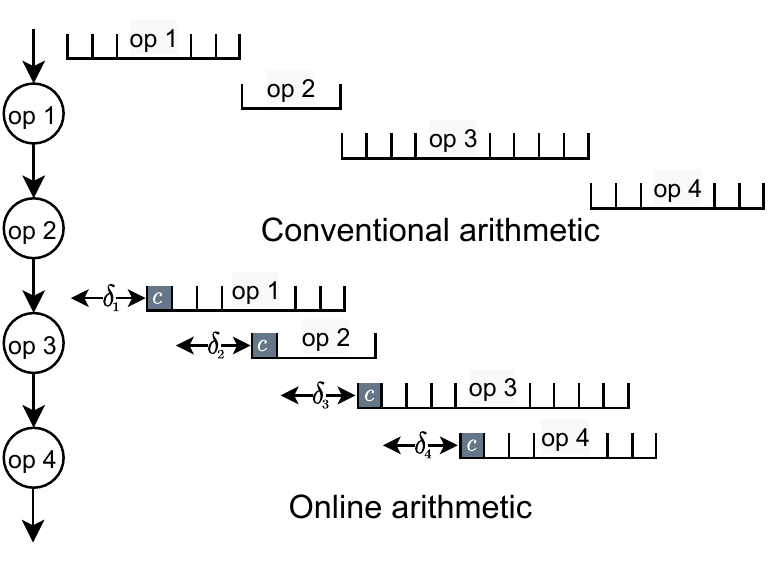}
	\end{center}
	\caption{Timing comparison of conventional and online arithmetic for a sequence of dependent operations.}
 	\label{fig: ppa}
\end{figure}

Several researchers have considered online arithmetic for implementing complex DSP algorithms on hardware to achieve high degree of parallelism \cite{tangtrakul1996signed, galli2001design, dormiani2005design}. Requirement of reduced interconnection bandwidth makes them suitable for adoption in the multi-module structures in both parallel and pipelined configurations where the interconnection bandwidth is constrained. The online delay $\delta$, during which the inputs are received, is independent of precision and, therefore, pipelining of the serially produced output digits is possible with the latency of $\delta$ in contrast to the digit-parallel algorithms where throughput is governed by the data dependency \cite{ercegovac2017left}. The timing difference between conventional and online arithmetic is such that for the sequence of dependent operations in the online arithmetic, the computation can be started as soon as the MSD of the result is generated from the preceding operation i.e., after $\delta$ + compute cycles. On the contrary, the conventional arithmetic operators must wait for the completion of previous computation. This phenomenon, assuming $\delta_i =3$ and compute cycle $c=1$, has been depicted in Fig.~\ref{fig: ppa}.

The output is computed on the basis of partial information about the inputs, therefore, the redundant number system is employed which allows a number to be represented in more than one way. Although, the cost per bit is increased, it results in an overall advantage, as the need for carry propagation is eliminated and cycle time of the operation becomes independent of bit-precision. We employ signed digit (SD) redundant number system where number representation is done in radix ($r$) form and each signed digit belongs to a set $\{-a,...,-1,0,1,...,a\}$ and $\frac{r}{2} \leq a < r$. For a digit set to be redundant, the digit set must fulfill the condition $2a+1 >r$ such that $a \leq r-1$. The amount of redundancy in the number system is represented by $\rho$ ($\rho = \frac{a}{r-1}$). A digit set is said to be redundant if  $\rho > \frac{1}{2}$, whereas if $\frac{1}{2}< \rho <1$, the digit set is called minimally redundant, while $\rho = 1$ and $\rho > 1$ constitutes to maximally and over-redundant digit sets respectively. For simplicity, all implementations in this study use radix-$2$ SD representation on a symmetric redundant digit set of $\{-1,0,1\}$. In cases where the conversion of SD to conventional number system is required, an efficient conversion technique named on-the-fly conversion (OTFC) is adopted \cite{ercegovac1987fly}. The process of OTFC does not require carry propagate adders, hence the computation is carried out without any additional delay. 

\subsection{Non-pipelined Online Multiplier}\label{sec: Olmult}
The online arithmetic algorithms including online adder and online multiplier use fractional numbers to make them compatible with all operations and to simplify the alignment of the operands, therefore, the weight of the operand's first digit is $r^{-1}$. At any $j^{th}$ iteration, a digit $x_j$ is represented by two single bits, $x_j^+$ and $x_j^-$, and their subtraction produces the value of the represented digit \eqref{eq: digit}, allowing a conventional number to be represented in several ways.
\begin{equation}\label{eq: digit}
		x_{j}= SUB (x_j^+,x_j^-)
	\end{equation}
	
The numerical value of the digit $x$ at iteration $j$ and $j+1$ is denoted as $x[j]$, and $x[j+1]$ respectively, and its corresponding online form is represented as:
\begin{equation}\label{eq: online_input}
    \begin{split}
 &x[j]= \sum_{i=1}^{j+\delta}x_{i}r^{-i}\\
 &x[j+1] = x[j]+x_{(j+1+\delta)} r^{-(j+1+\delta)}
    \end{split}
\end{equation}

\begin{figure}[!ht]
	\begin{center}
\includegraphics[width=1.0\textwidth]{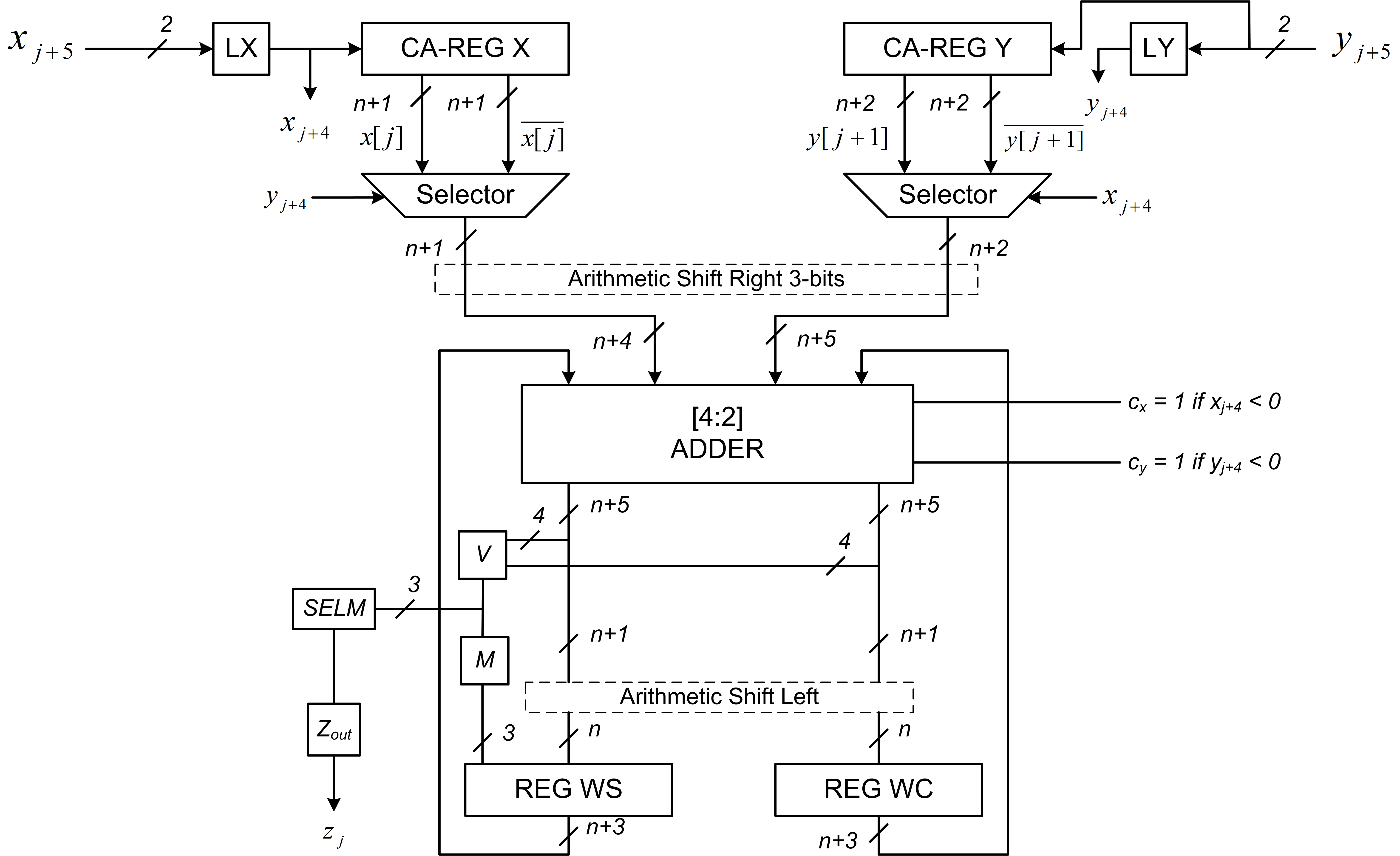}
	\end{center}
	\caption{Non-pipelined radix-2 online multiplier \cite{ercegovac2004digital}. Online delay $\delta$=3 and $t$=2.}
 	\label{fig: R2Mult}
\end{figure}

The non-pipelined radix-$2$ online multiplier with $n$ precision as shown in Fig.~\ref{fig: R2Mult}, presented in \cite{ercegovac2004digital}, has an online delay $\delta =3$ and the selection function requires $t=2$ fractional bits along with $2$ integer bits (\emph{ibs}) to select the output. The description of each module in the figure is presented in section \ref{Implementation}. The input operands $x$ and $y$ in the signed digit redundant representation are computed to produce the product digit $z$ ranging from $(-1,1)$ from the symmetric signed digit set $\{-a,...,a\}$. The operands and the resulting product digit at iteration $j$ are given as:
\begin{equation}
x[j] = \sum_{i=1}^{j+\delta}x_{i}r^{-i},\;\;\;\;\; \\
y[j] = \sum_{i=1}^{j+\delta}y_{i}r^{-i},\;\;\;\;\; \\
z[j] = \sum_{i=1}^{j}z_{i}r^{-i},
\end{equation}
where the subscripts denote the digit index and the iteration index is indicated by square brackets.

The algorithm executes for $n+\delta$ iterations, during which one digit of input operands $x_j$ and $y_j$ is introduced per iteration, except for the last $\delta$ cycles where the input digits are set to zero. Similarly the output digit $z_j$ is produced at each iteration after $\delta$ cycles, whereas, in the first $\delta$ iterations, the result for the output digit remains zero. The fundamental part of the online algorithms is the development of recurrence on the internal state and defining a selection function with selection constants to produce the result digit. Both these methodologies and the algorithm of the online multiplier have been detailed in the ensuing section.

\subsubsection{Residual and its Recurrence}
Method for developing online algorithms including addition, multiplication, and division have been presented in Chapter $9$ of \cite{ercegovac2004digital}. For completeness, we present the derivations of residual and recurrence in the online multiplication in the following. At each iteration $j$, a SD input is received such that the entrance of one of the operands (in this case $y$) is one clock cycle prior to the other. The SD input is converted to two's complement representation in digit serial manner using on-the-fly conversion/append (CA) function as: $x[j] = CA(x[j-1],x_{j+4})$ and $y[j] = CA(y[j-1],y_{j+5})$. An output is produced on the basis of only partial information of the inputs, therefore, an error bound must be defined as follows:

\begin{equation}\label{eq: err_bound}
\lvert x[j] \cdot y[j] -z[j] \rvert < r^{-j}
 \end{equation}

The above relation is subjected to a transformation function to develop the recurrence having primitive functions only, which is then scaled by a factor to have a bound on the error after the computation of $j$ digits. The corresponding scaled residual is given by:
\begin{equation}
    w[j] = r^j (x[j]\cdot y[j]-z[j])
\end{equation}
The residual can be deduced to obtain the recurrence $w[j+1]$:
\begin{equation}
\begin{split}
w[j+1] & = rw[j] +(x[j] \cdot y_{j+1+\delta} + y[j+1] \cdot x_{j+1+\delta} )r^{-\delta} \\
& \;\;\;\;- z_{j+1}
\end{split}
\end{equation}
This can be decomposed into:
\begin{equation}\label{eq: recurrence}
    \begin{split}
        &v[j] = rw[j] +(x[j]\cdot y_{j+1+\delta}+ y[j+1]\cdot x_{j+1+\delta} )r^{-\delta} \\
&w[j+1] = v[j] - z_{j+1},
    \end{split}
\end{equation}
or,
\begin{equation}
\begin{split}
&v[j] = rw[j] + H_1\\
&w[j+1] = v[j] + H_2(z_{j+1})    
\end{split}
\end{equation}
resulting in, 
\begin{equation}
\begin{split}
    &H1 = (x[j]\cdot y_{j+1+\delta} + y[j+1]\cdot x_{j+1+\delta})r^{-\delta}\\
    &H2 = -z_{j+1},
\end{split}
\end{equation}

so that $H_1$ is independent of $z_{j+1}$. 

For $r=2$ and $\delta = 3$, $v[j]$ in \eqref{eq: recurrence} can be rewritten as:
\begin{equation}\label{eq: recurrence2}
v[j] = 2w[j] +(x[j]\cdot y_{j+4}+ y[j+1] \cdot x_{j+4} )2^{-3} 
\end{equation}
The multiplication of terms with $2^{-3}$ in \eqref{eq: recurrence2} is carried out using arithmetic right shift by $3$. 
As can be observed in Fig.~\ref{fig: R2Mult}, the residual, $w[j]$, in the redundant carry-save form actually has a $2's$ complement representation and is represented by the vectors $WS[j]$ and $WC[j]$.

Next step is to determine the bounds of $w[j+1]$ in terms of $H_1$ and $H_2$. This is given as:
\begin{equation}\label{eq: bound1}
    \overline{\omega} = r\overline{\omega}+\text{max}(H_1) + H_2(a)
\end{equation}
resulting in 
\begin{equation}\label{eq: omegaplus}
    \overline{\omega} = -\frac{\text{max}(H_1)+H_2(a)}{r-1}
\end{equation}
Likewise,
\begin{equation}\label{eq: omegaminus}
    \underline{\omega} = -\frac{\text{min}(H_1)+H_2(-a)}{r-1}
\end{equation}
For the case of serial-serial multiplier, max($H_1$) and min($H_1$) results in $2ar^{-\delta}$ and $-2ar^{-\delta}$ respectively, whereas $H_2(a)$ and $H_2(-a)$ are $-a$ and $a$ respectively. Substituting values of $H_1$ and $H_2$ in relations \ref{eq: omegaplus} and \ref{eq: omegaminus}, yields following results
\begin{equation}\label{eq: omegaplus1}
    \overline{\omega} = -\frac{2ar^{-\delta} - a}{r-1}
\end{equation}
Likewise,
\begin{equation}\label{eq: omegaminus1}
    \underline{\omega} = \frac{2ar^{-\delta}-a}{r-1}
\end{equation}

\subsubsection{Selection Function with Selection Constants}\label{sec: t}
Two methods have been suggested for the selection of output digit in \cite{ercegovac2004digital}, one of which uses selection constants, while the other method is based on rounding of the residual which is used for higher radix ($r>4$). In this research, we employ radix-$2$, therefore, the selection function is implemented using selection constants. The output digit $z_{j+1} = q$ where $q_i= -a,-a+1, \ldots, a$ depends upon the selection intervals of $v[j]$ is selected using a selection function such that the residual $w[j+1]$ remains bounded. Only $t$ most significant fractional bits along with integer bits of $v[j]$ are used from the result generated by the $[4:2]$ adder in carry-sum pair ($WS$ and $WC$) to give its estimate, $\widehat{v}[j]$. 

The output digit $z_{j+1}$ is produced using the selection function in a way that $w[j+1]$ remains bounded according to relations \ref{eq: omegaplus} and \ref{eq: omegaminus}. In the method with selection constants it is described by the selection constants ${m_k}$ such that
\begin{equation}
    z_{j+1} = k \;\;\; \text{if} \;\;\;m_k \leq \hat{v}[j] < m_{k+1} 
 \end{equation}

Here $\hat{v}[j]$ is an estimate of $v[j]$, computed by truncating $v[j]$ to $t$ fractional bits.
To produce a correct selection function, the selection constants must satisfy
\begin{equation}
    \text{max}(\widehat{L}_k) \leq m_k \leq \text{min}(\widehat{U}_{k})
\end{equation}
where [$\widehat{L}_k,\widehat{U}_{k}$] is the selection interval of the estimate $\hat{v}[j]$. 
The selection intervals [${L}_k,{U}_{k}$] are obtained from relation, and then the intervals are restricted for $\hat{v}[j]$. 
\begin{equation}\label{eq: omegaU}
    \overline{\omega} = U_k + H_2(k) \;\;\;\;\; \underline{\omega} = L_k + H_2(k)
\end{equation}
An error is introduced due to truncation and using estimate $\hat{v}[j]$, given as
\begin{equation}
    e_{min} \leq v[j] - \hat{v}[j] \leq e_{max}
\end{equation}
For carry-save representation, $e_{max} = 2^{-t+1} - ulp$ and $e_{min} = 0$, which when substituted in eq.~(\ref{eq: omegaU}) for $\widehat{L}_k$ and $\widehat{U}_k$, and using carry-save representation, for $w[j]$ and $v[j]$, results in the following
\begin{equation}
    \begin{split}
        &\widehat{U}_{k} = {\lfloor \rho(1-2r^{-\delta})+k-2^{-t}\rfloor}_t \\
        &\widehat{L}_{k} = {\lceil - \rho(1-2r^{-\delta})+k\rceil}_t
    \end{split}
\end{equation}

To determine $t$ and $\delta$, we use the relation $\widehat{U}_{k-1} - \widehat{L}_{k} \geq 0$. The corresponding expression is given as
\begin{equation}\label{eq: t and delta}
        {\lfloor \rho(1-2r^{-\delta})+k-1-2^{-t}\rfloor}_t - \widehat{L}_{k} = {\lceil - \rho(1-2r^{-\delta})+k\rceil}_t \geq 0
    \end{equation}
Since $\rho=1$, and radix $r=2$ is known,  we substitute one variable, either $t$ or $\delta$ to obtain the value of another. The objective is to obtain small values of both the variables. Starting from minimum values the relation is checked for satisfaction. For serial-serial multiplier, the relation is satisfied for $\delta=3$ and $t=1$. 
The selection constants $m_k$'s are obtained from
\begin{equation}
    \widehat{L}_k \leq m_k \leq \widehat{U}_{k-1}
\end{equation}
which results in $m_0 = -2^{-1}$ and $m_1 = 2^{-1}$.

The range of $\hat{v}[j]$ is given by:
\begin{equation}\label{eq range of vj for serial-serial}
{\lfloor r\underline{\omega}+\text{min}(H_1) - e_{max} \rfloor \leq \hat{v}[j] \leq \lceil  r\overline{\omega}+ \text{max}(H_1)+\lvert e_{min}\rvert \rceil}_t    
\end{equation}
Substituting corresponding values we obtain $-2 \leq \hat{v}[j] \leq \frac{7}{4}$. This is used to define the selection function \textit{SELM} as shown in relation (\ref{eq: SELM}).


	\begin{equation}\label{eq: SELM}
		z_{j+1} =SELM (\widehat{v}[j]) = \begin{cases}
		\;\;\,1 &\text{if $1/2 \leq \widehat{v}[j] \leq 7/4$}\\
		\;\;\,0 &\text{if $-1/2 \leq \widehat{v}[j] \leq 1/4$}\\
		-1 &\text{if $-2 \leq \widehat{v}[j] \leq -3/4$ }
		\end{cases} 
	\end{equation}
	
The product digit $z_{j+1}$ uses similar coding as \eqref{eq: digit} and the corresponding selection function is shown in Table.~\ref{tbl: LUT}. The estimate of the residual is calculated in the $V$ block and the calculation of the updated  residual $w[j+1]$, which requires subtraction of $z_{j+1}$ from $v[j]$, is carried out by the $M$ block. The subtraction is performed using a Boolean expression rather than explicit subtraction \cite{dormiani2005design}, as shown later in section \ref{sec: M_module}.

\subsubsection{Algorithm}
The conventional algorithm has two steps; (1) initialization: having execution length equal to $\delta$, during which the input digits are collected and no output is generated, (2) recurrence: which executes for $n$ iterations, producing one output digit in each iteration. The pseudocode of the non-pipelined radix-2 serial-serial online multiplier is shown in Algorithm \ref{alg:algorithm1}.  
   \begin{algorithm}
		  	\begin{algorithmic}[1]
		  	\State {Initialize:\newline $x[-3]=y[-3]=w[-3] = 0$}
            \For{j=$-3,-2,-1$}
                \State$x[j+1] \leftarrow CA\left(x[j], x_{j+4}\right); 
                \newline \phantom{x}\hspace{2.2ex} y[j+1] \leftarrow CA\left(y[j], y_{j+4}\right);$
                \State{$v[\jmath]=2 w[\jmath]+\left(x[\jmath] y_{j+4}+y[j+1] x_{j+4}\right) 2^{-3}$}
                \State{$w[\jmath+1] \leftarrow v[j]$}
            \EndFor \newline
             \State{Recurrence:}
             \For{$j=0 \ldots n+\delta$}
             \State$x[j+1] \leftarrow CA\left(x[j], x_{j+4}\right); 
                \newline \phantom{x}\hspace{2.2ex} y[j+1] \leftarrow CA\left(y[j], y_{j+4}\right);$
                \State{$v[j]=2w[j]+\left(x[j] y_{j+4}+y[j+1] x_{j+4}\right) 2^{-3}$}
                \State{$z_{j+1}=SELM(\widehat{v[j]})$}
                \State{$w[j+1] \leftarrow v[j]-z_{j+1}$}
                \State{$Z_{\text {out}} \leftarrow z_{j+1}$}
             \EndFor \newline

\end{algorithmic}
\caption{Online Multiplication}
\label{alg:algorithm1}
  \end{algorithm}
  
\subsection{Non-pipelined Serial-Parallel Online Multiplier}
\begin{figure}[!ht]
	\begin{center}
\includegraphics[width=0.7\textwidth]{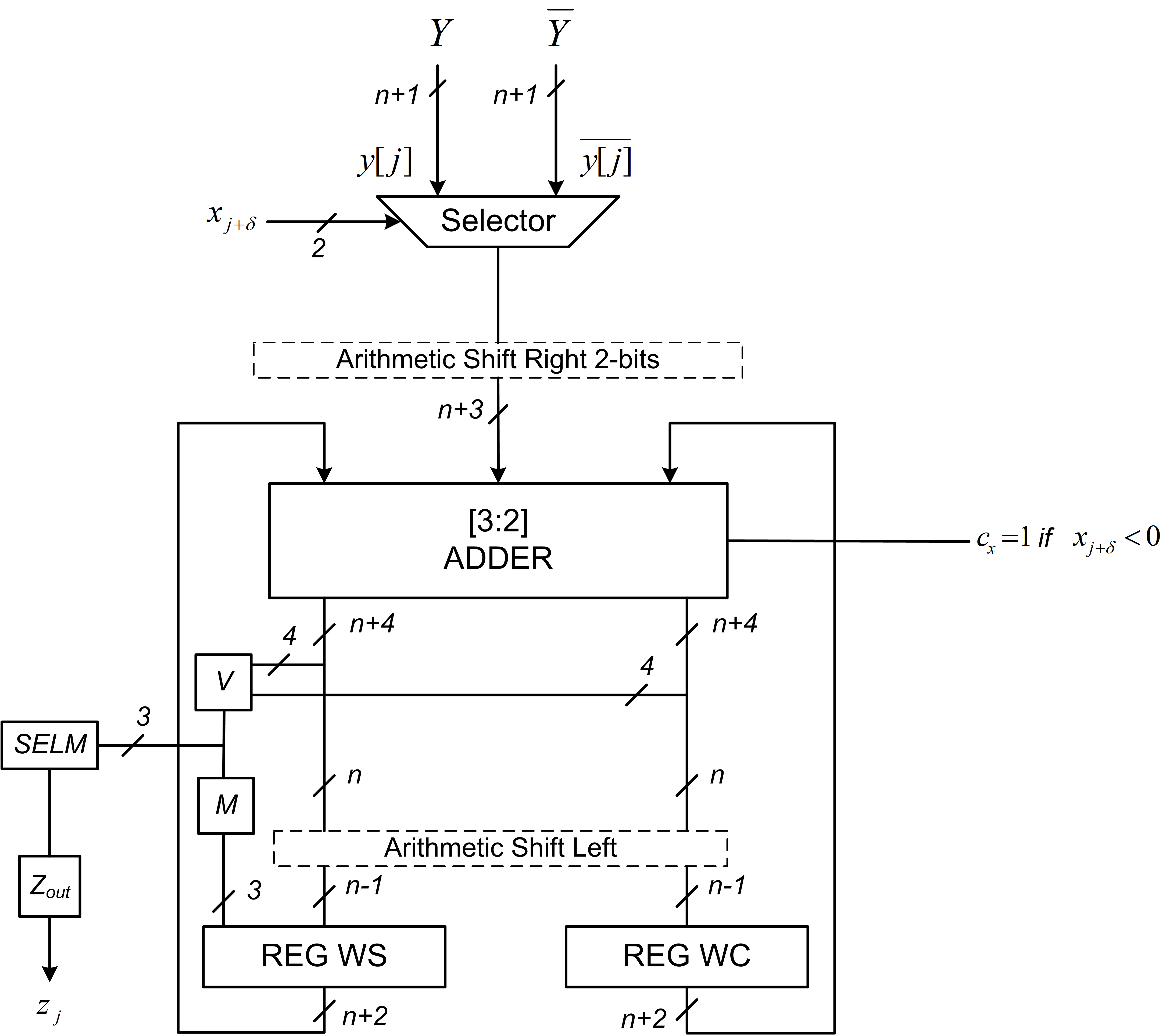}
	\end{center}
	\caption{Non-pipelined radix-2 serial-parallel online multiplier. Online delay $\delta$=2 and $t$=2.}
 	\label{fig: R2Mult_SP}
\end{figure}

For the serial-parallel multiplier, one of the operands enters in serial MSDF manner and the other is a constant and is available in parallel at the implementation time. The radix-$2$ non-pipelined serial-parallel online multiplier depicted in Fig.~\ref{fig: R2Mult_SP}, has an online delay of $2$. Its selection function requires $t=2$ fractional and $1$ integer bit.  The derivation of online serial-parallel multiplier has been presented in \cite{galli2001design}. It follows same steps as the online multiplier with both operands in serial, therefore, only the results of the derivations are presented in this section. 

The online operands (input $x$ and output $z$) in each cycle are represented as \eqref{eq: online_input}, and the constant is represented as:
\begin{equation}
    Y[j] = Y = -y_0 \;.\; r^{0}+ \sum_{i=1}^{n}y_{i}r^{-i}
\end{equation}
\subsubsection{Residual and its Recurrence}
The error bound at each computation step is given as:
\begin{equation}
    \lvert x[j] \cdot Y -z[j]\rvert < r^{-j}
\end{equation}

The scaled residual at step $j$ is given as:
\begin{equation}
    w[j] = r^j (x[j]\cdot Y-z[j])
\end{equation}
Recurrence in the function $w[j+1]$ is obtained as:
\begin{equation}
\begin{split}
w[j+1] & = rw[j] +(x_{j+\delta}\cdot Y )r^{-\delta}- z_{j+1},
\end{split}
\end{equation}

The bounds of $w[j+1]$ are determined similarly as (\ref{eq: bound1}), (\ref{eq: omegaplus}), and (\ref{eq: omegaminus}). However, for serial-parallel online multiplier max($H_1$) and min($H_1$) are $ar^{-\delta}$ and $-ar^{-\delta}$ respectively. 

The resulting $\overline{\omega}$ and $\underline{\omega}$ are given as:
\begin{equation}\label{eq: omegaplus2}
    \overline{\omega} = -\frac{ar^{-\delta} - a}{r-1},
\end{equation}
and,
\begin{equation}\label{eq: omegaminus2}
    \underline{\omega} = \frac{ar^{-\delta}-a}{r-1}
\end{equation}
\subsubsection{Selection Function with Selection Constants}\label{sec: t1}
The selection function for serial-parallel online multiplier is obtained in a similar manner as for the serial-serial online multiplier derived in \ref{sec: t}. However, the corresponding values of $H_1$ and $H_2$ are substituted in relations to obtain $\hat{U_k}$ and $\hat{L_k}$ as follows: 
\begin{equation}
    \begin{split}
        &\widehat{U}_{k} = {\lfloor \rho(1-r^{-\delta})+k-2^{-t}\rfloor}_t \\
        &\widehat{L}_{k} = {\lceil - \rho(1-r^{-\delta})+k\rceil}_t
    \end{split}
\end{equation}

Similarly the values of $t$ and $\delta$ are determined using following relation:
\begin{equation}\label{eq: t and delta for SP}
        {\lfloor \rho(1-r^{-\delta})+k-1-2^{-t}\rfloor}_t - \widehat{L}_{k} = {\lceil - \rho(1-r^{-\delta})+k\rceil}_t \geq 0
    \end{equation}
    
Relation (\ref{eq: t and delta for SP}) yields $t=2$ and $\delta=2$ for the online serial-parallel multiplier. They are then used to determine the range of $\hat{v}[j]$ using (\ref{eq range of vj for serial-serial}), which in turn yields similar values as of serial-serial multiplier. The selection constants for serial-parallel multiplier are also similar, therefore, the same selection function (\ref{eq: SELM}) is utilized for serial-parallel online multiplier with $2$ integer bits. 
\subsubsection{Algorithm}
Similar to serial-serial online multiplier, the algorithm has two steps, (1) initialization: having execution length equal to $\delta$ during which, the input digits are collected and no output is generated, (2) recurrence: which executes for $n$ iterations, producing one output digit in each iteration. The pseudocode of the non-pipelined radix-2 serial-parallel online multiplier is shown in Algorithm \ref{alg:algorithm_SP}.  
   \begin{algorithm}
		  	\begin{algorithmic}[1]
		  	\State {Initialize:\newline $x[-2]=w[-2] = 0$}
            \For{j=$-2,-1$}
                \State{$v[\jmath]=2 w[\jmath]+\left(x_{j+2} \cdot Y] \right) 2^{-2}$}
                \State{$w[\jmath+1] \leftarrow v[j]$}
            \EndFor \newline
             \State{Recurrence:}
             \For{$j=0 \ldots n+\delta$}
             
                \State{$v[\jmath]=2 w[\jmath]+\left(x_{j+2} \cdot Y] \right) 2^{-2}$}
                \State{$z_{j+1}=SELM(\widehat{v[j]})$}
                \State{$w[j+1] \leftarrow v[j]-z_{j+1}$}
                \State{$Z_{\text {out}} \leftarrow z_{j+1}$}
             \EndFor \newline
\end{algorithmic}
\caption{Serial-Parallel Online Multiplication}
\label{alg:algorithm_SP}
  \end{algorithm}

\section{Proposed Pipelined Online Multiplier}\label{sec: proposed}

In the non-pipelined design of online multiplier, the working precision of $n$ bits is constant and all digit slices remain active during all iterations. However, in \cite{ercegovac2020}, sources of reduction of active slices have been presented which include gradual use of the input digits and reduction of working precision to $p<n$. The proposed pipelined design is a $2$D implementation in which the inactive modules are not implemented which results in savings of both dynamic and static power.

\subsection{Precision Reduction} \label{sec: p}
\begin{figure}[!ht]
	\begin{center}
\includegraphics[viewport=10 0 215 75,scale=.75]{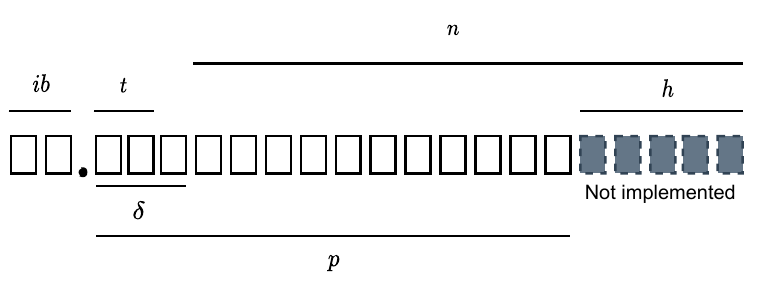}
	\end{center}
	\caption{Bit slice reduction \cite{ercegovac2004digital}. Online delay $\delta$=3 and $t$=2.}
 	\label{fig: p_analysis}
\end{figure}

Since the output digit of the online algorithm is based on a selection function which utilizes a few most significant bits of the residual comprised of an integer and $t$ fractional bits, as discussed in section \ref{sec: t}, to obtain the residual's estimate denoted by $\hat{v}$. Therefore, it is possible to achieve $n$ bits accuracy by implementing only $p$ ($p<n$) bit slices and ignoring a few least significant $h$ bit slices as shown in Fig.~\ref{fig: p_analysis}.

For $j \leq p$ iterations $j+\delta$ modules are active in $j^{th}$ recurrence step; whereas, for $j>p$, the availability of only $p$ modules introduces an error due to truncation. The algorithm's convergence can be assured if the $t$ bits in the selection function are not affected due to the truncation error. The optimal number of $p$ varies according the type of adder used in the recurrence equation, the number of ignored bit slices $h$, and the initial delay $\delta$. For a valid selection in an online multiplier with $[4:2]$ adder, and using $t$ fractional bits $p-2h+\delta \geq t$. Since $p+h =n+\delta$ we obtain relation \eqref{eq: pcalc} as suggested in \cite{ercegovac2004digital}:
\begin{equation}\label{eq: pcalc}
    p = \left \lceil \frac{2n +\delta +t}{3}  \right \rceil
\end{equation}
Due to the gradual increase in the precision of the incoming digits, the signal activity is not constant and increases gradually in each iteration. Furthermore, if $p<n$ slices are implemented for the given multiplier, the signal activity begins to decrease after $p$ iterations due the truncation error which affects $(j-2)^{th}$ result bit and is shifted one bit towards left due to the left shift operation in the recurrence. Overall, the error propagates to $3$ bit slices, therefore, the $3$ least significant bit slices can be turned off in the subsequent stage of the pipeline yielding a low-power design. 

Based on these properties, $8, 16, 24$ and $32$ bit low-power designs of pipelined serial-serial multiplier have been compared with pipelined serial-serial online multiplier with full working precision in \cite{usman2021}. According to relation \eqref{eq: pcalc}, the $n$ precision result can be obtained by employing $7, 12, 18$ and $23$ modules for $8, 16, 24$ and $32$ bit designs respectively. The proposed low-power design has been implemented as a two-dimensional pipeline array, where the bit widths of the registers (\emph{CA-Reg}, \emph{Reg WS}, \emph{Reg WC}), adder, and selector are increased till $p^{th}$ iteration and then decreased till $n+\delta$ iteration. 

\subsection{Pipelined Online Multiplier}
\begin{figure}[!ht]

\centering
\includegraphics[width=1.2\textwidth, angle =90 ]{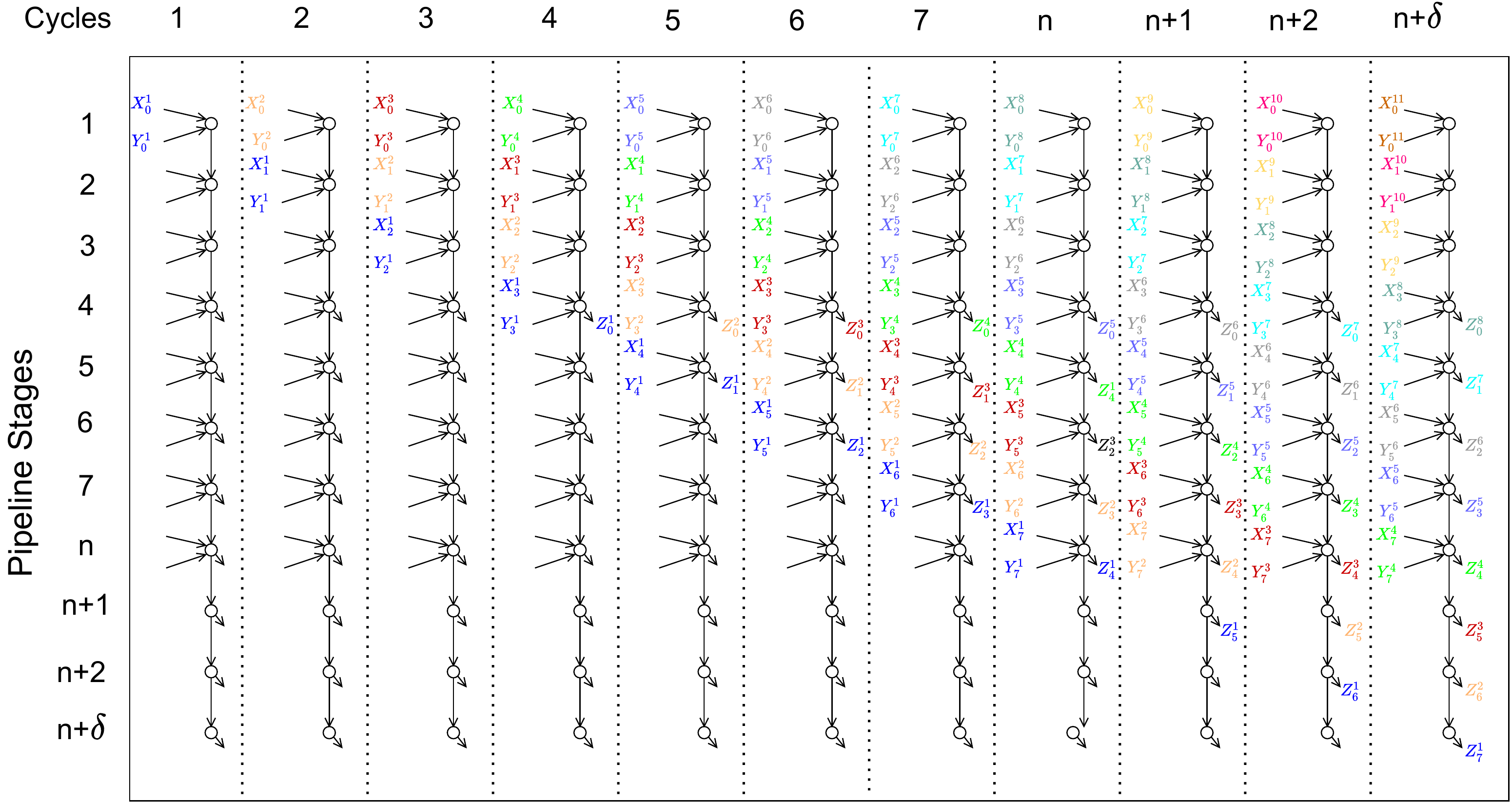}

	\caption{Multiplication of $K$ vectors with operands $X$ and $Y$, producing output vector
	$Z$ in a digit-level pipelined online multiplier, assuming $\delta = 3$ and compute cycle = $1$. Superscript denote the vector index $K \in \{1,2 \ldots k\}$, subscripts denote the bit index $n \in \{0,1, \ldots 7\}$. In practice, the output is latched (not shown in figure) and is received with a delay of $1$ clock cycle, i.e., complete result of $Z^0$ is obtained in $n+\delta+1$ cycle followed by $Z^1$ in $n+\delta+2$ cycle and so forth.}
 	\label{fig: Pipeline_Function}
\end{figure}

The non-pipelined online multiplier has its throughput limited by its latency because it produces one vector in $n+\delta+1$ cycles. In applications where large number of multiplications have to be performed, this limitation on the throughput may not be acceptable. Therefore, to process large number of operations, it is suitable to unfold and pipeline the multiplier. For $n$-bit precision, $n+\delta$ stages of the multiplier are unfolded and pipelined. It takes $n+\delta$ cycles to fill the pipeline, and once the pipeline is in steady state, the multiplier produces $n$-bit output vector in each cycle. This phenomenon has been depicted in Fig.~\ref{fig: Pipeline_Function}, for $8$ bit wide $K$ vectors of $X$ and $Y$. The cycle time for the pipelined online multiplier is the same as the non-pipelined online multiplier and is also independent of bit precision. This drastically improves the throughput of the network. As discussed in section \ref{sec: p}, the input bit precision is increased gradually and $p<n$ modules are sufficient to produce $n$-bit precision result, only the required number of modules can be activated upto $p^{th}$ iteration and after truncation in $(p+1)^{th}$ iteration, the modules can be turned off using some switching mechanism according to the error profile. In a pipelined scheme, however, the inactive modules are not implemented, hence no dynamic/static power is consumed. 


\subsubsection{Algorithm}
We present the algorithm in three steps \cite{usman2021}, including, (1) initialization: having execution length equal to $\delta$, during which the input digits are collected and no output is generated, (2) recurrence: which executes for $n-\delta$ iterations, producing one output digit in each iteration. (3) last $\delta$ cycles: having execution length equal to $\delta$, during which the input digits are $zero$ and output is generated in each iteration. The pseudocode of the non-pipelined radix-2 online multiplier presented in \cite{ercegovac2004digital}, with initialization and recurrence loops, has been modified to have three loops as shown in the Algorithm \ref{alg:algorithm1Proposed}, the corresponding block diagrams were presented in \cite{usman2021}.
  
   \begin{algorithm}
		  	\begin{algorithmic}[1]
		  	\State {Initialize:\newline $x[-3]=y[-3]=w[-3] = 0$}
            \For{j=$-3,-2,-1$}
                \State$x[j+1] \leftarrow CA\left(x[j], x_{j+4}\right); 
                \newline \phantom{x}\hspace{2ex} y[j+1] \leftarrow CA(y[j], y_{j+4});$
                \State{$v[\jmath]=2 w[\jmath]+\left(x[\jmath] y_{j+4}+y[j+1] x_{j+4}\right) 2^{-3}$}
                \State{$w[\jmath+1] \leftarrow v[j]$}
            \EndFor \newline
             \State{Recurrence:}
             \For{$j=0 \ldots n-\delta-1$}
             \State$x[j+1] \leftarrow CA\left(x[j], x_{j+4}\right); 
                \newline \phantom{x}\hspace{2ex} y[j+1] \leftarrow CA(y[j], y_{j+4});$
                \State{$v[j]=2w[j]+\left(x[j] y_{j+4}+y[j+1] x_{j+4}\right) 2^{-3}$}
                \State{$z_{j+1}=SELM(\widehat{v[j]})$}
                \State{$w[j+1] \leftarrow v[j]-z_{j+1}$}
                \State{$Z_{\text {out}} \leftarrow z_{j+1}$}
             \EndFor \newline
            \State{Last $\delta$ cycles:}
             \For{$j=n-\delta \ldots n-1$}
                \State {$x[n-\delta \ldots n-1]=y[n-\delta \ldots n-1]=0$}
                \State{$v[j]=2w[j]$}
                \State{$z_{j+1}=SELM(\widehat{v[j]})$}
                \State{$w[j+1] \leftarrow v[j]-z_{j+1}$}
                \State{$Z_{\text {out}} \leftarrow z_{j+1}$}
             \EndFor
\end{algorithmic}
\caption{Proposed Online Multiplication}
\label{alg:algorithm1Proposed}
  \end{algorithm}

    



\begin{figure}[!ht]
	\begin{center}
\includegraphics[viewport=15 8 180 110,scale=1.0]{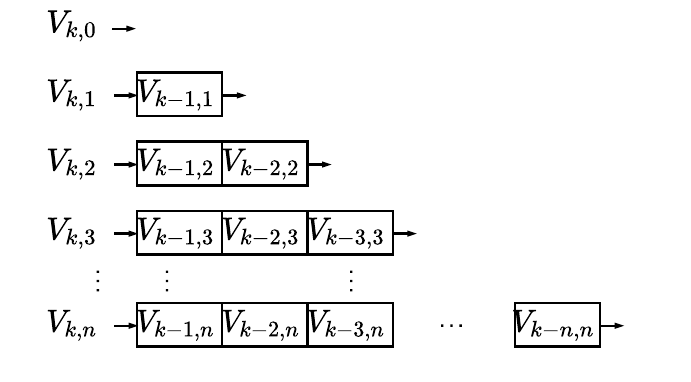}
	\end{center}
	\caption{Stair-case input shifter array \cite{huang2001fpga}.}
 	\label{fig: array}
\end{figure}

\begin{figure*}[ht]
	\begin{center}
\includegraphics[viewport=20 3 415 265,scale=0.8]{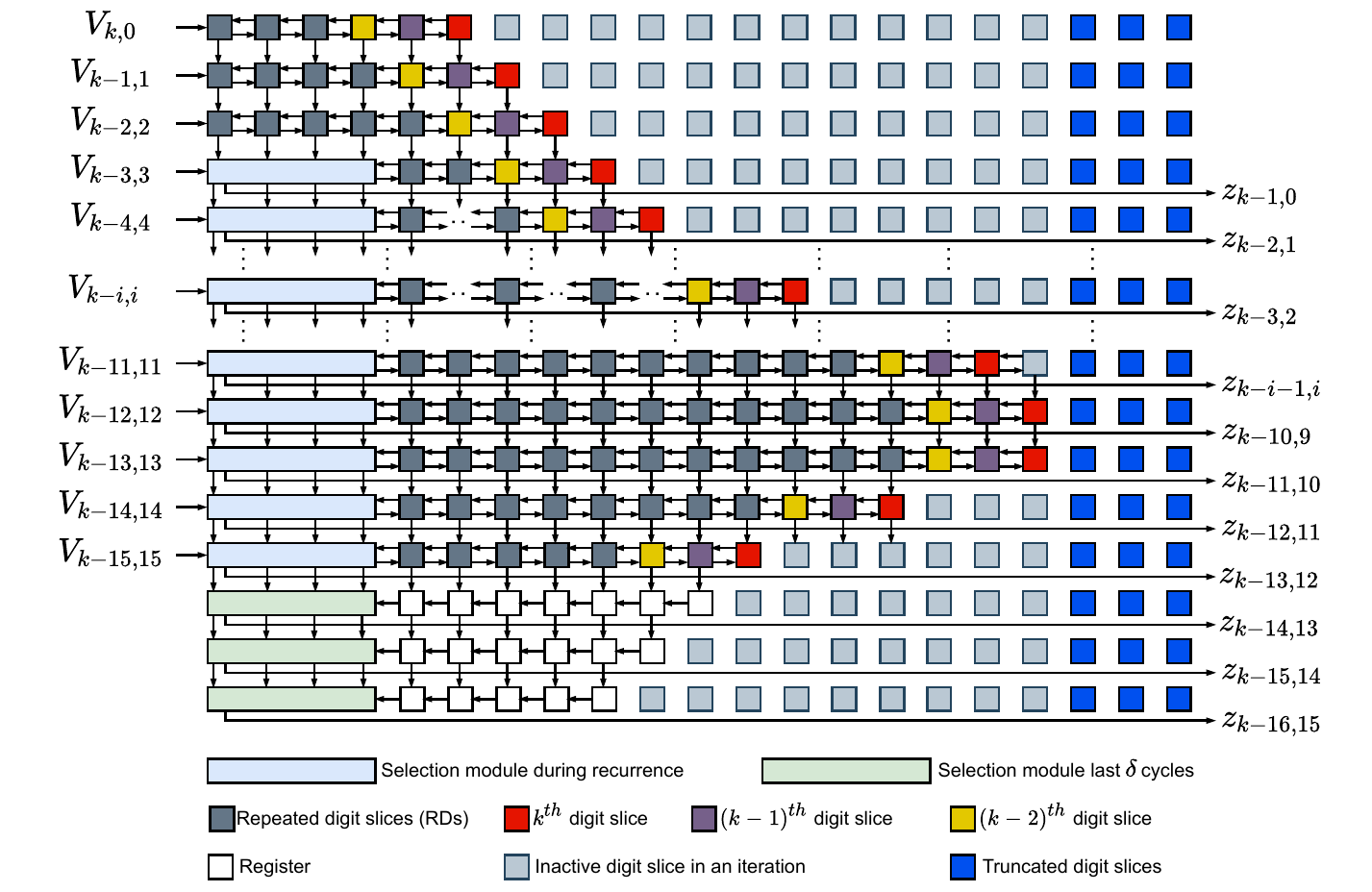}
	\end{center}
	\caption{Signal activity of radix-2 online multiplication algorithm with truncated working precision of $p$ with $\delta=3$, $2$ integer bits (\emph{ib}) and \emph{t} $=2$. Different colors of the digit slice refer to their distinct structure as discussed in section \ref{sec: adder_init}. SEL block evaluates three bits to compute the output and is therefore larger than the rest of digit slices \cite{usmanthesis}. }
 	\label{fig: POL}
\end{figure*}

A $16$-bit pipelined scheme which is a two-dimensional array structure with $16$ stages has been depicted in Fig.~\ref{fig: POL}. The digit selection module in the most significant place is instantiated after initialization steps to generate an output digit and the residual signals are transferred vertically to the subsequent linear array instead of left shifting as in the conventional implementation. The input vectors are arranged in a stair-case manner to match the pipeline online flow using a stair-case shifter array shown in Fig.~\ref{fig: array}. This simply adds a delay in the $i^{th}$ digit of a vector using an $i$-bit shift register \cite{huang2001fpga}. The details of each digit slice has been presented in the forthcoming section. 

\subsection{Implementation Details}\label{Implementation}
In the proposed design, each digit slice has been fine tuned in order to reduce signal activities, area utilization and power consumption. Accordingly, only useful modules are instantiated in each step of the three sub-loops. We present the details of each module and the corresponding digit slice structure in the following.

\subsubsection{Initialization}
\begin{figure}[!ht]
	\begin{center}
\includegraphics[viewport=8 8 325 175,scale=0.6]{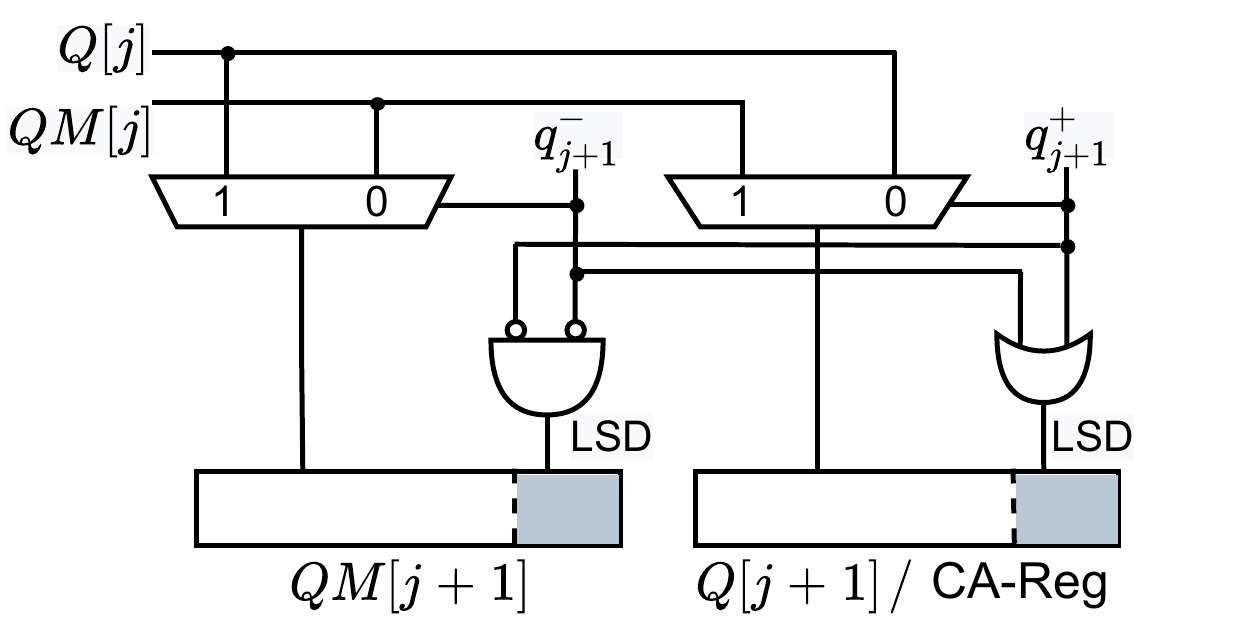}
	\end{center}
	\caption{Digit slice of on-the-fly converter.}
 	\label{fig: OTFC}
\end{figure}

During initialization, the algorithm executes for $\delta$ cycles to accumulate sufficient input digits to produce the first output. Since no output digit is produced during initialization, the modules to generate the output digit are not implemented. The digit slices in the initialization consists of OTFC units and selectors. While the presence of adders either half, full, or their combination depends on the position of the digit slice. The detail of each unit in the initialization stage is as follows:
\paragraph{On-the-fly Conversion}
The redundant SD inputs are required in the conventional form during the recurrence step $j$, which are obtained without any additional delay using the OTFC module; proposed in \cite{ercegovac1987fly}. Two OTFC units are instantiated for the two operands during initialization and recurrence, each composed of two $2-to-1$ multiplexers, $2$-input \emph{OR} and \emph{AND} gates and two registers to store \emph{Q} and \emph{QM} = $Q-1$ as shown in Fig.~\ref{fig: OTFC}. In each iteration, a new incoming digit is appended in the least-significant digit of either \emph{Q} and \emph{QM} registers depending on the value of $q_{j+1}$, increasing its width by one bit upto \emph{n+ib}. Two integer bits are initialized as `$00$' or `$11$' representing `$0$' and `$-1$' for the first positive or negative fractional bit respectively. The conversion/append (\emph{CA-Reg}) registers correspond to the \emph{Q[j+1]} register of the OTFC unit. According to the online multiplier's algorithm, the computation requires advance availability of one of the operands (in this case operand \emph{y}), therefore, the bit width for `$y$' OTFC unit is one bit longer than `$x$' in all iterations.  

\begin{figure}[!ht]
	\begin{center}
\includegraphics[viewport=10 2 220 80,scale=1.0]{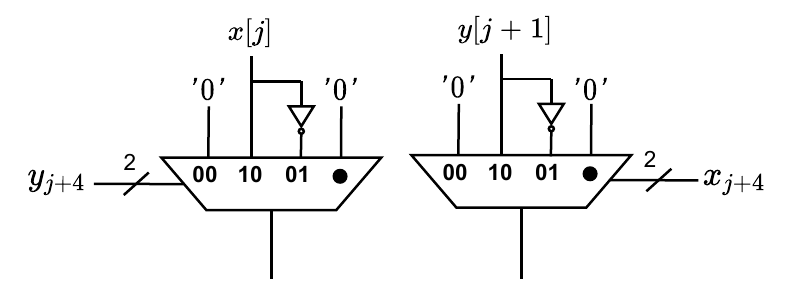}
	\end{center}
	\caption{Digit slice of selector unit.}
 	\label{fig: Sel}
\end{figure}

\paragraph{Selector}
Multiplication is performed by \emph{selector}, which is a $4-to-1$ multiplexer as shown in Fig.~\ref{fig: Sel}. Since it receives the inputs from the \emph{CA-Reg} registers, the width of its inputs also increases upto \emph{$p^{th}$} iteration, and then begins to decrease. As there are no inputs in the last $\delta$ cycles, the selector module is instantiated during initialization and recurrence stages only. In each iteration, the signed digit selector can take values from $1, -1$ or $0$, encoded as `$10$',`$01$' and `$00$', for which the selector outputs \emph{x.y}, $\overline{\emph{x.y}}$, or $0$ respectively. 

\begin{figure}[!ht]
    \centering
    \includegraphics[width=0.5\textwidth]{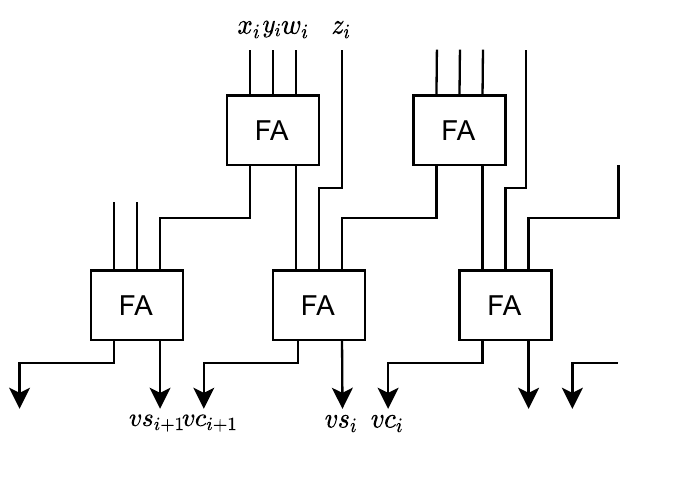}
    \caption{[4:2] carry-save adder using two stages of full adders \cite{ercegovac2004digital}.}
    \label{fig:4-2 adder}
\end{figure}

\paragraph{Adder}\label{sec: adder_init}
A $[4:2]$ carry-save adder (CSA) is employed to perform the addition of the input operands and the residual. The functionality of this adder is obtained by utilizing two full adders. Intermediate sums and carries are denoted by $VS$ and $VC$ respectively, whereas the final output vectors of sum and carry are denoted as \emph{vs} and \emph{vc} respectively. The requirement is to add the two carry-save operands in \emph{WS} and \emph{WC}, with the two conventional operands ($x[j]\cdot y_{j+4}$) and ($y[j+1]\cdot x_{j+4}$). The reduction is implemented by two carry-save adders as shown in Fig.~\ref{fig:4-2 adder}. The $[4:2]$ adder has fixed delay of two full adders, which is significantly smaller than the carry-propagate adders.

At any given iteration \emph{j}, the number of bits in \emph{x} and \emph{y} are \emph{k+ib+$\delta$} and \emph{k+ib+$\delta$+1} respectively, where \emph{k} are the number of fractional bits in a given iteration. The length of residual registers \emph{WS} and \emph{WC} is \emph{k+ib+$\delta$-1} during initialization. 

\begin{figure*}[ht]
	\begin{center}
	\includegraphics[width=1.0\textwidth]{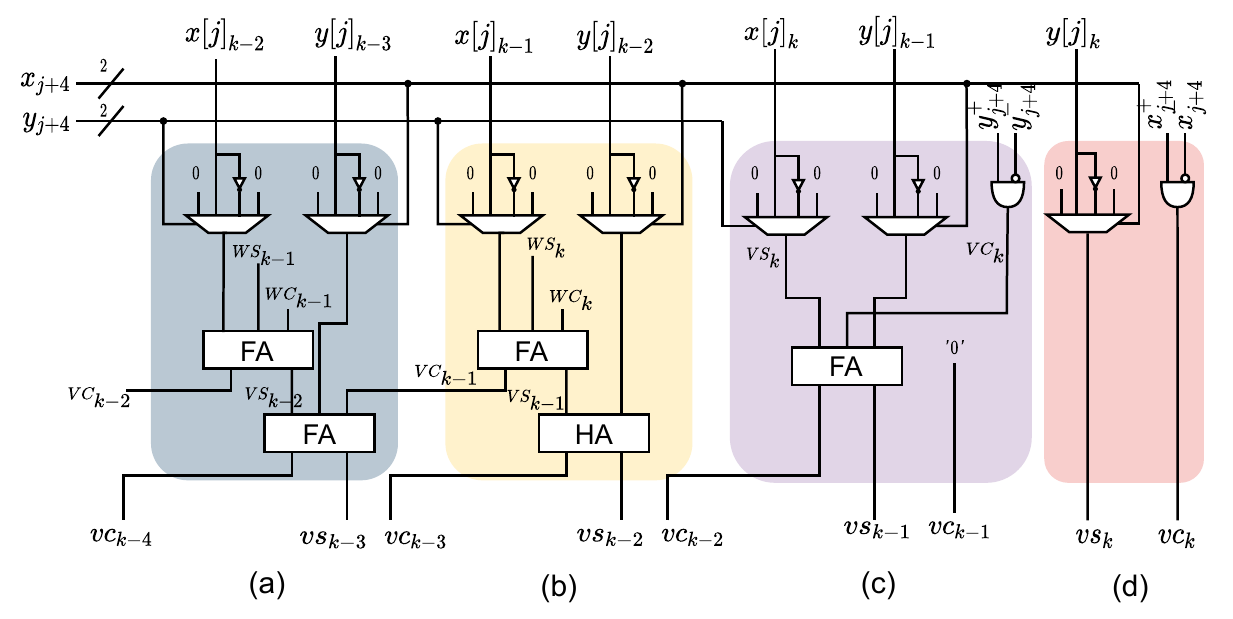}
	\end{center}
	\caption{Internal structure for the least significant and repeated digit slices.}
 	\label{fig: LSBs}
\end{figure*}

For the low-power implementation, a distinct structure of adder is specified according to the bit position, as shown in different colors in Fig.~\ref{fig: POL}. For the two's complement representation, multiplication with `$-1$' can be performed by negating the input bits and adding a logical $1$ to the \emph{unit in the last place (ulp)}, therefore, the least significant bit $vc_k$ of the $vc[j]$ vector is designated for $c_x$ ($c_x = x_{j+4}^+ \cdot \overline{x_{j+4}^-}$). The corresponding digit slice to obtain the least significant bits of $vc$ and $vs$ is the red colored slice $k$ from the Fig.~\ref{fig: POL} and its internal circuit is depicted in Fig.~\ref{fig: LSBs} (d). It has no adders because the length of vector $y[j]$ is largest and there are no digits to be added, therefore, $y_k$ and $c_x$ are simply copied to $vs_k$ and $vc_k$ respectively.

For the same reason of achieving correct result of multiplication of a vector by `$-1$' in the two's complement, the least significant bit $VC_k$ of the $VC[j]$ is accounted for $c_y$ ($c_y = y_{j+4}^+ \cdot \overline{y_{j+4}^-}$). Since the length of vector $x[j]$ is one digit smaller than $y[j]$, $c_y$ is present in the $(k-1)^{th}$ digit slice. This is the purple colored slice in Fig.~\ref{fig: POL} and its internal circuit is depicted in  Fig.~\ref{fig: LSBs} (c). The length of the recurrence registers is one bit smaller than that of $x[j]$, therefore a single full adder is employed to add the three input digits $x[j]_m$, $y[j]_m$, and $c_y$. Furthermore, absence of an adder in the $k^{th}$ place accounts for no output carry, therefore, a permanent `$0$' is placed at the $vc_{k-1}$ position. Due to a single full adder in $(k-1)^{th}$ position, there are no intermediate sum or carry digits, instead, a final sum $vs_{k-1}$ and a carry $vc_{k-2}$ is produced. This implies that in the $(k-2)^{th}$ digit slice, a full adder in the first stage and a half adder in the second stage is sufficient to produce the outputs. This slice is shown in yellow color in Fig.~\ref{fig: POL} while its logic is shown in Fig.~\ref{fig: LSBs} (b). The $(k-2)^{th}$ digit slice however, generates both intermediate and final carry digits to the higher digit slice, therefore, the $(k-3)^{th}$ digit slice is composed of two full adders. First full adder evaluates the sum of $x[j]$, $WS[j]$, and $WC[j]$ and produces an intermediate carry and sum vector named as $VC[j]$ and $VS[j]$ respectively. The second full adder evaluates the sum of $VS[j]$, $VC[j]$, and $y[j+1]$ to produce final sum and carry, expressed as, $vs[j]$ and $vc[j]$ respectively, and are collectively represented as $v[j]$ shown in Eq.~\ref{eq: 2vj}. This grey colored digit slice from Fig.~\ref{fig: POL} is implemented using the logic shown in Fig.~\ref{fig: LSBs} (a). It is named as repeated digit slice (RD) as the same digit slice is repeated $k+ib$ times during initialization. 

\paragraph{Residual Calculation}
In the initialization phase, the residual for the next iteration ($2w[j+1]$) corresponds to the left shifting of $vs$ and $vc$ vectors. The most significant \emph{ibs} of both vectors (i.e., $vs_{-1}$ and $vc_{-1}$) are discarded and the vectors are left shifted by a simple re-wiring. The updated residual ($2w[j+1]$) is shown in relation \eqref{eq: 2wj}. 

\begin{equation}\label{eq: 2wj}
2w[j+1] \Big\rvert \begin{array}{llll}
vs_{0} \; vs_{1} \cdot vs_{2}\; vs_{3}\;  vs_{4}\; vs_{5} \ldots \\
vc_{0} \; vc_{1} \cdot vc_{2}\; vc_{3}\;  vc_{4}\; vc_{5}\;  \ldots
\end{array}
\end{equation}

\subsubsection{Recurrence}
After accumulating sufficient number of input digits to generate the output, the algorithm advances to the recurrence stage. SEL digit slice is instantiated to generate the output.  The computation of next residual ($2w[j+1]$) involves $M$ block which subtracts the output digit $z[j+1]$ from $\hat{v}[j]$. Similar structures of the OTFC and the selector modules shown in Fig.~\ref{fig: OTFC} and Fig.~\ref{fig: Sel}, respectively, are utilized for the entire recurrence stage with bit widths corresponding to the signal activity pattern.  
The circuit diagram for the blue colored SEL digit slice from Fig.~\ref{fig: POL} has been  depicted in Fig.~\ref{fig: MSBs}. 

Residual's estimate calculation, selection of the output digit, and subtraction of the output digit from residual are performed by distinct modules present in the SEL digit slice, details of each of these modules are briefed below:
\begin{figure*}[ht]
	\begin{center}
\includegraphics[viewport=5 12 515 285,scale=0.65]{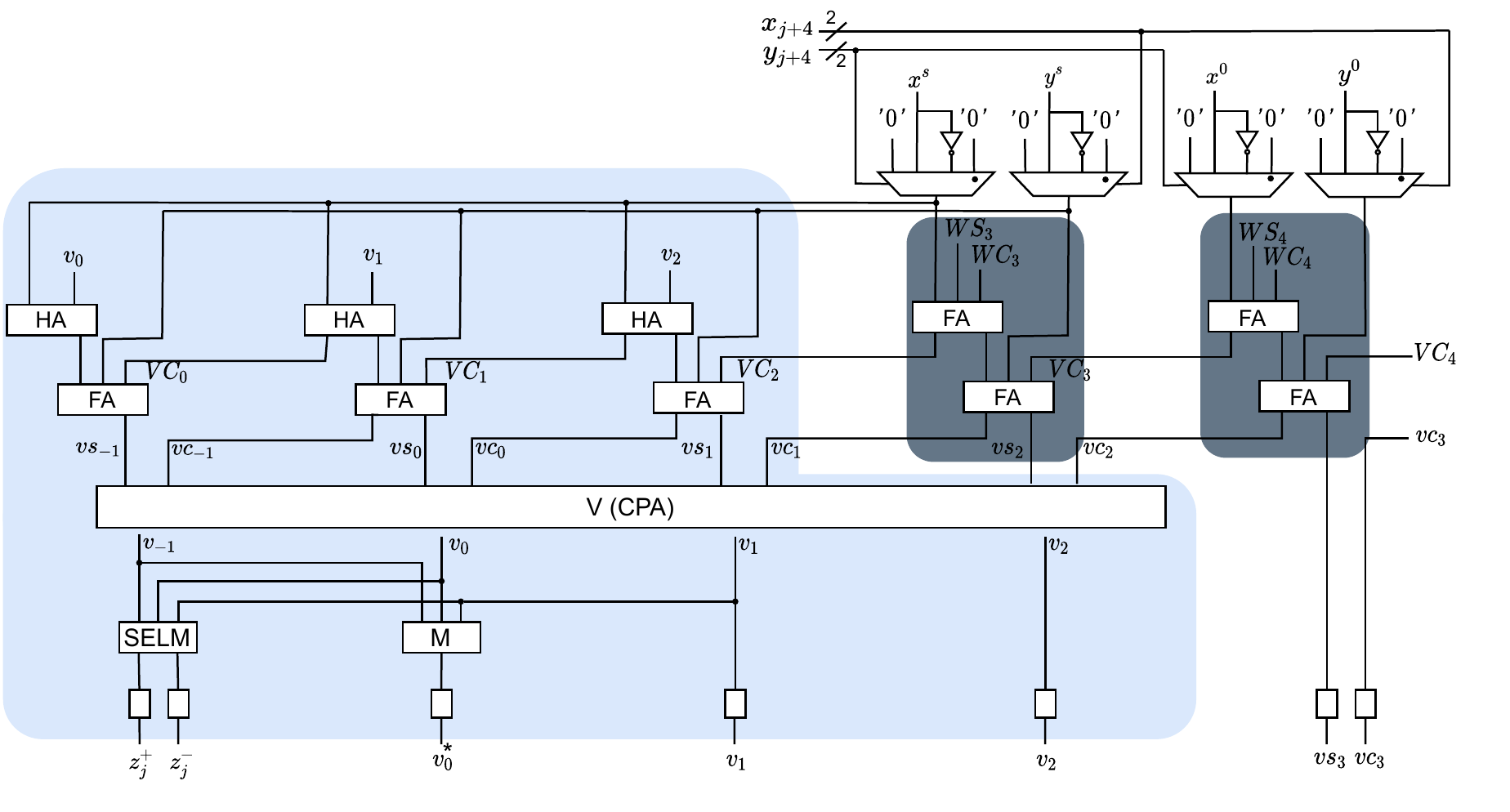}
	\end{center}
	\caption{Logic for the SEL digit slice during recurrence. The three MSBs are composed of a combination of half and full adders in contrast to two full adders in the repeated digit slices.}
 	\label{fig: MSBs}
\end{figure*}

\paragraph{V Block}
The output is based on the estimate ($\hat{v}[j]$) of the residual (\emph{v[j]}) and is evaluated in the $V$ block. It is a carry propagation adder that performs the addition of $t$ most significant fractional bits and the integer bits of $v[j]$ (represented by $vs[j]$ and $vc[j]$ vectors shown in Eq.\eqref{eq: 2vj}) to generate the estimate of the residual ($\hat{v}$) as shown in Eq.~\eqref{eq: v_est}. 

\begin{equation}\label{eq: 2vj}
v[j] \Big\rvert \begin{array}{llll}
vs_{-1} \; vs_{0} \cdot vs_{1}\; vs_{2}\;  vs_{3}\; vs_{4} \ldots \\
vc_{-1} \; vc_{0} \cdot vc_{1}\; vc_{2}\;  vc_{3}\; vc_{4}  \ldots
\end{array}
\end{equation}
\begin{equation}\label{eq: v_est}
\hat{v} \Big\rvert \begin{array}{llll}
v_{-1} \; v_{0} \cdot v_{1}\; v_{2} \ldots
\end{array}
\end{equation}

\paragraph{SELM Module} \label{sec: SELM}
The result of the $V$ block is subjected to the \emph{SELM} module for selecting the corresponding output from a look-up table shown in Table.~\ref{tbl: LUT}. In case of radix-2 online multiplier, the least significant estimate bit $v_2$ is not used and the three MSBs i.e., two \emph{ibs} ($v_{-1}$ and $v_{0}$) and one fractional bit ($v_{1}$) are sufficient to select the output $z_{j+1}$. 
\renewcommand{\arraystretch}{1.0}
\begin{table}
\caption{Selection function for serial-serial and serial-parallel radix-2 multiplier.}
\begin{center}
\resizebox{0.3\textwidth}{!}{
\begin{tabular}{ccc} \hline \hline
$\hat{v}$ & $v_{-1} v_{0} . v_{1}$ & $z_{j+1}$ \\ \hline \hline
\multicolumn{1}{c|}{3/2} & \multicolumn{1}{c|}{01.1} & \multicolumn{1}{c}{1} \\ 
\multicolumn{1}{c|}{1} & \multicolumn{1}{c|}{01.0} & \multicolumn{1}{c}{1} \\ 
\multicolumn{1}{c|}{1/2} & \multicolumn{1}{c|}{00.1} & \multicolumn{1}{c}{1} \\ 
\multicolumn{1}{c|}{0} & \multicolumn{1}{c|}{00.0} & \multicolumn{1}{c}{0} \\ 
\multicolumn{1}{c|}{-1/2} & \multicolumn{1}{c|}{11.1} & \multicolumn{1}{c}{0} \\ 
\multicolumn{1}{c|}{-1} & \multicolumn{1}{c|}{11.0} & \multicolumn{1}{c}{-1} \\ 
\multicolumn{1}{c|}{-3/2} & \multicolumn{1}{c|}{10.1} & \multicolumn{1}{c}{-1} \\ 
\multicolumn{1}{c|}{-2} & \multicolumn{1}{c|}{10.0} & \multicolumn{1}{c}{-1} \\ \hline \hline
\end{tabular}}
\end{center}
\label{tbl: LUT}
\end{table}

\paragraph{M Block}\label{sec: M_module}
It performs the subtraction of $z_{j+1}$ from the residual's estimate ($\hat{v}$) to produce $2w[j+1]$. The subtraction to obtain $v_0^*$ is performed using the following Boolean expression \cite{dormiani2005design}:
\begin{equation}\label{eq: bool}
    v_0^*= v_0 \; XOR \; \rvert p_{j+1}\rvert \\
\end{equation}

\paragraph{Adders}
The RDs and the least significant digit slices for adder are similar to the initialization stage, and for $k$ bit precision, $k+ib-3$ number of RDs are instantiated in a certain iteration. The length of the vector $vc$ after being subjected to $V$ block is reduced by $3$ bits (refer to Eq.~\eqref{eq: 2wjup}). Therefore, in the $3$ most significant bit slices, which accounts for the SEL block, instead of two stages of full adders, a half adder is employed in the first stage and a full adder is employed in the second stage. The multiplication of the terms $x[j]\cdot y_{j+4}$ and $y[j+1]\cdot x_{j+4}$ in the recurrence equation with $2^{-3}$ in both initialization and recurrence stages corresponds to the sign extension of the MSBs, which is done by performing $3$ bit arithmetic right shift operation without any additional cost. 

\paragraph{Residual}\label{sec: Res_Rec}
The MSB of $\hat{v}[j]$ i.e., $\hat{v}_{-1}$ is discarded and the remaining $3$ bits are vertically transferred to $vs$ vector consequently resulting in an updated residual as shown in relation \eqref{eq: 2wjup}. In this manner the left shifting of the residual ($2w[j]$) is carried out. 

\begin{equation}\label{eq: 2wjup}
2w[j+1] \Big\rvert \begin{array}{llll}
v_{0} \; v_{1} \cdot v_{2}\; v_{3}\;  v_{4}\; v_{5} \ldots \\
\, \phantom{xxxxxxx} v_{3}\;  v_{4}\; v_{5}  \ldots
\end{array}
\end{equation}
\subsubsection{Last $\delta$ cycles}
\begin{figure}[!ht]
	\begin{center}
\includegraphics[viewport=10 6 220 125,scale=1.0]{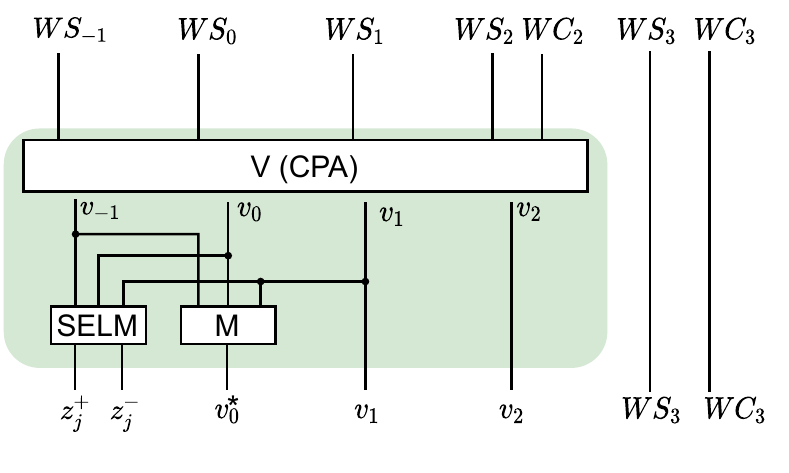}
	\end{center}
	\caption{Logic for the MSB in last $\delta$ cycles. The SEL digit slice is simplified to a carry propagation adder, SELM, and M block; taking the shifted residuals as input and producing an output in each cycle.}
 	\label{fig: PR_MSB}
\end{figure}

The remaining output digits are obtained in the last $\delta$ iterations which produces one output digit in each cycle. All the inputs are utilized in the initialization and recurrence stages in the non-pipelined online multiplier, and three $0s$ are applied as an input for last $\delta$ iterations. However, in the proposed low-power design, all unused modules are eliminated and therefore, the OTFC, selector, and $[4:2]$ adders are not implemented. The residual containing two vectors \emph{WS} and \emph{WC} are subjected to the \emph{V}, \emph{M}, and \emph{SELM} modules to perform their respective tasks and generate the output digit. The digit slice used during the last $\delta$ iterations has been depicted in Fig.~\ref{fig: PR_MSB}. 

\subsection{Pipelined Serial-Parallel Multiplier}
With one operand available in parallel, the circuitry for serial-parallel multiplier is simplified. As shown in Fig.~\ref{fig: R2Mult_SP}, the multiplier does not have OTFC modules and requires [$3:2$] adder i.e., only one stage of full adder in each bit slice. The range of input and output is different from serial-serial, and requires one integer bit which also simplifies the selection function. The truncation strategy for serial-parallel multiplier has not been adopted, rather the entire $n$-bit operand is utilized during all iterations. However, as shown in algorithm \ref{alg:algorithm_PSP}, the algorithm has been divided into three sections namely initialization, recurrence, and last $\delta$ cycles. 

   \begin{algorithm}[!ht]
		  	\begin{algorithmic}[1]
		  	\State {Initialize:\newline $x[-2]=w[-2] = 0$}
            \For{j=$-2,-1$}
                \State{$v[\jmath]=2 w[\jmath]+\left(x_{j+2} \cdot Y] \right) 2^{-2}$}
                \State{$w[\jmath+1] \leftarrow v[j]$}
            \EndFor \newline
             \State{Recurrence:}
             \For{$j=0 \ldots n+\delta$}
             
                \State{$v[\jmath]=2 w[\jmath]+\left(x_{j+2} \cdot Y] \right) 2^{-2}$}
                \State{$z_{j+1}=SELM(\widehat{v[j]})$}
                \State{$w[j+1] \leftarrow v[j]-z_{j+1}$}
                \State{$Z_{\text {out}} \leftarrow z_{j+1}$}
             \EndFor \newline
            \State{Last $\delta$ cycles:}
             \For{$j=n-\delta \ldots n-1$}
                \State{$v[j]=2w[j]$}
                \State{$z_{j+1}=SELM(\widehat{v[j]})$}
                \State{$w[j+1] \leftarrow v[j]-z_{j+1}$}
                \State{$Z_{\text {out}} \leftarrow z_{j+1}$}
             \EndFor
\end{algorithmic}
\caption{Proposed Pipelined Serial-Parallel Online Multiplication}
\label{alg:algorithm_PSP}
  \end{algorithm}

\begin{figure}[hbt!]
    \centering
    \begin {subfigure}[]{0.45\textwidth}
    \centering
    \includegraphics[width=\textwidth]{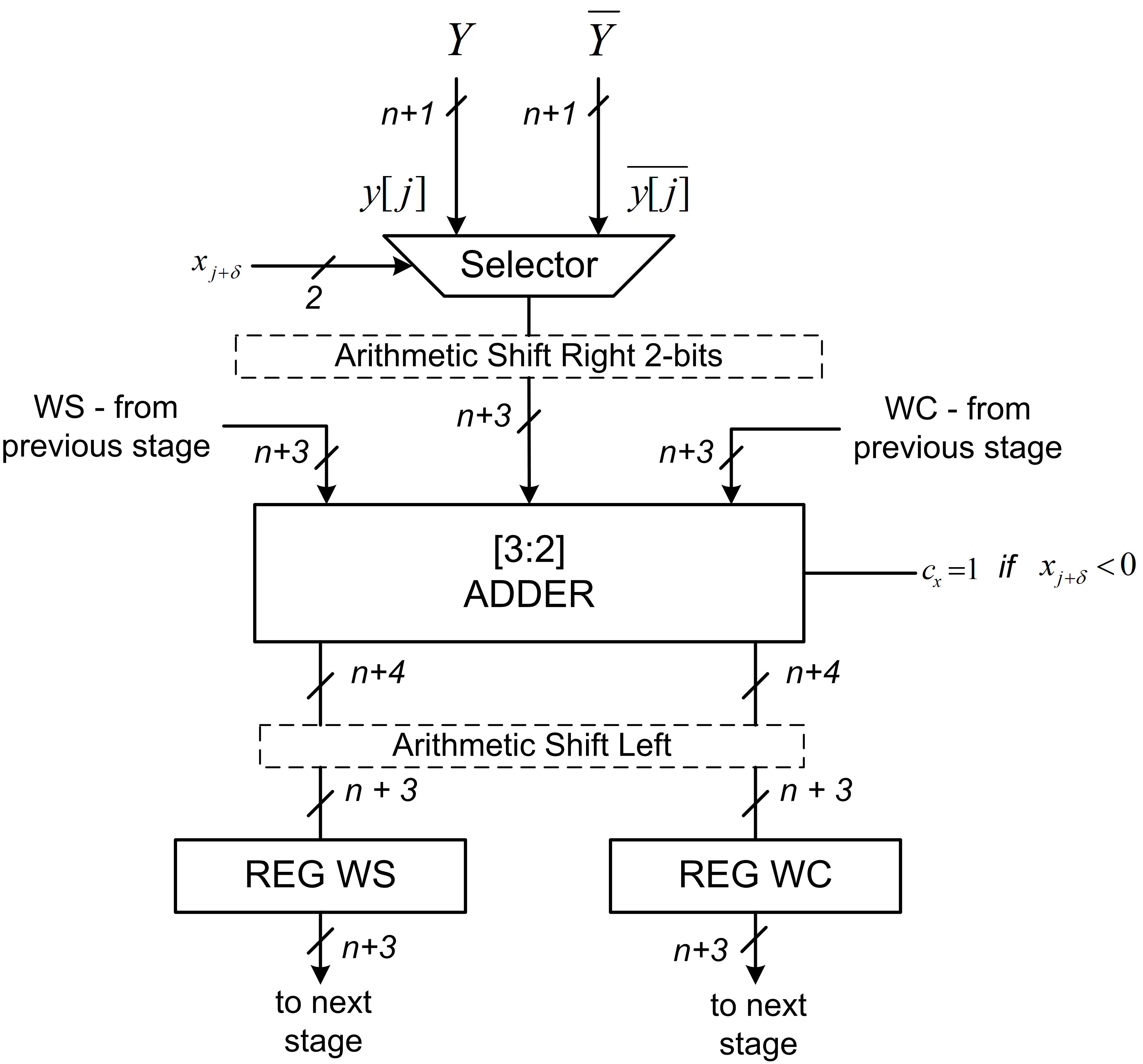}
    \caption{Initialization stage: inputs are received and number bit-slices increase in each cycle. No output is produced, therefore, $V$, $M$ and $SELM$ modules are not instantiated.}
    \label{fig: Initialization}
    \end{subfigure}
    \hfill
        \begin {subfigure}[]{0.5\textwidth}
    \centering
    \includegraphics[width=\textwidth]{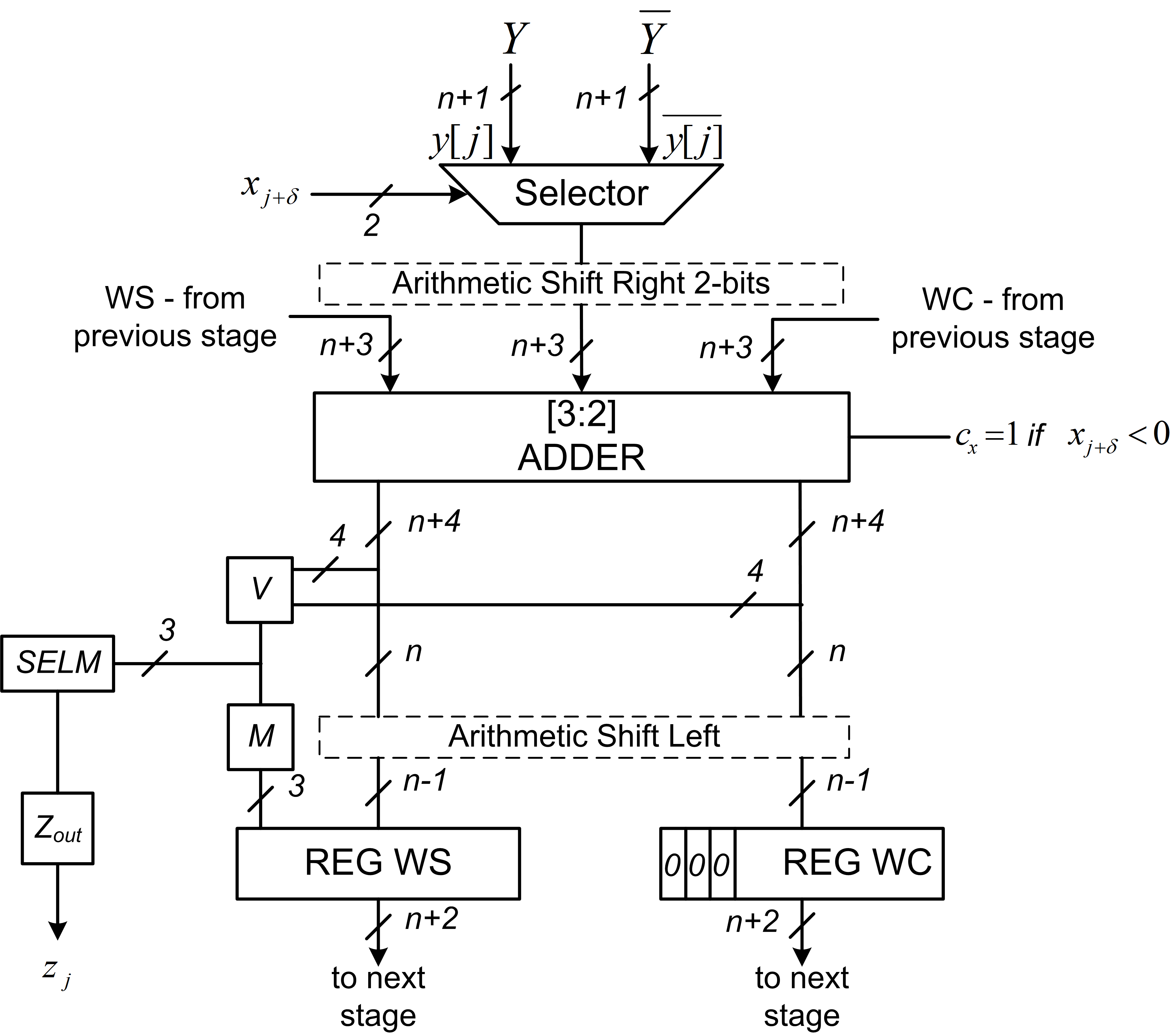}
    \caption{Recurrence stage: all modules are functional and follow gradual activation/deactivation of bit-slices.}
    \label{fig: Recurrence}
    \end{subfigure}
    \hfill
    
    \begin {subfigure}[]{0.45\textwidth}
    \centering

    \includegraphics[width=\textwidth]{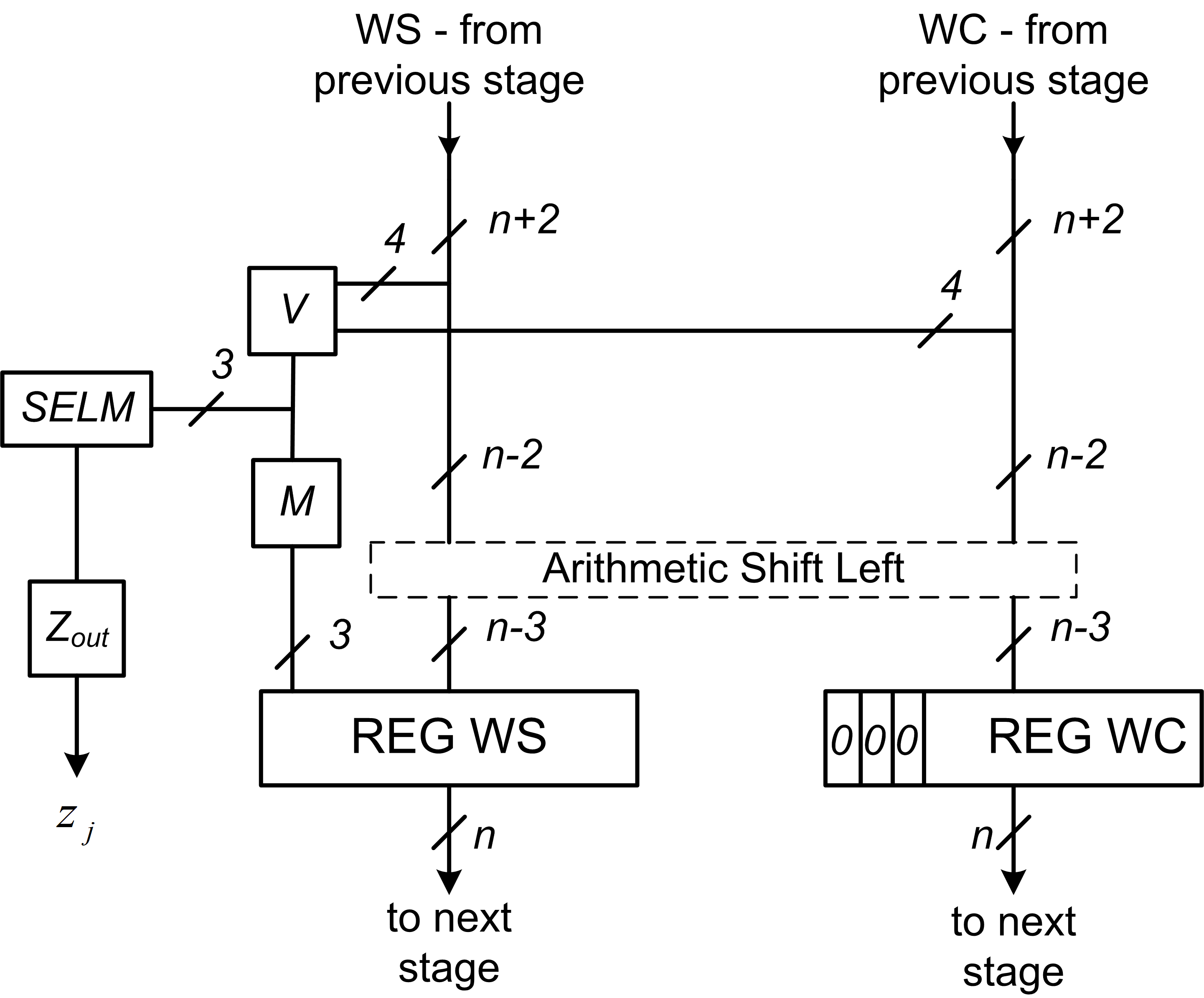}
        \caption{Last $\delta$ cycles: input is $0$, therefore, the input reception modules are removed from the design.}
    \label{fig: last delta}
    \end{subfigure}
    \caption{Proposed stages of pipelined online multiplier}
    \label{fig: proposed_block}
    \end{figure}

The block diagram corresponding to each step of the algorithm is shown in Fig.~\ref{fig: proposed_block}. It can be observed that in each stage, only the corresponding circuits are instantiated to reduce area utilization and power consumption. For instance, in the initialization step shown in Fig.\ref{fig: Initialization}, the modules for output generation are not instantiated. Likewise, in the last $\delta$ cycles, where the input is $0$, the modules for input are not instantiated as shown in Fig.~\ref{fig: last delta}. During recurrence however, an input is received and an output is produced in each cycle, the corresponding block diagram is depicted in Fig.~\ref{fig: Recurrence}. 

To select the output digit, the selection function takes $t=2$ fractional bits and $2$ integer bits from the estimate computed by $V$ block. The selection function is same as of serial-serial online multiplier shown in Table.~\ref{tbl: LUT}.


\section{Results}\label{sec: Results}
\subsection{Emperical Analysis}
We use Mentor Graphics ModelSim to verify the correctness of logical operation of the Verilog codes developed for the proposed and contemporary multipliers. A case study to show the effect of reduction in working precision has been presented for $16$-bit precision serial-serial multiplier with the following operands:
\begin{equation*}
    \begin{split}
& x= 00.110\overline{1}0\overline{1}\overline{1}011\overline{1}0\overline{1}100 \\
& y = 00.\overline{1}1\overline{1}100\overline{1}101\overline{1}11\overline{1}0\overline{1} 
    \end{split}
\end{equation*}
The numerical value of \emph{x} and \emph{y} is $0.66644287109375$ and $-0.31562805175781$ respectively. The actual product in conventional form is $-0.2103480650112033$ and the calculated product from online multiplier is $-0.2103424072265625$. The difference between the actual and calculated product is $5.657784640789032\times10^{-6}$ which is well under the error bound of the last iteration which is $2^{-16} = 1.52587890625\times10^{-5}$. Not only the final result, but the result of the online multiplier in each cycle is within the respective error bound according to relation \eqref{eq: err_bound}. In the event of variable precision requirement, the computation can be stopped upon reaching the desired precision, resulting in an accurate result upto that precision; unlike approximate circuits in which the result is not accurate. 

For reduced working precision for $n = 16$, $p = 13$ was evaluated from relation \eqref{eq: pcalc}. The digit slices are gradually increased according to the increasing precision of the inputs until $p-\delta$ cycles.
Truncation is applied in $p-\delta+1$ cycle that introduces an error in positions $p$, $p-1$ and $p-2$ which propagates to the left for residual calculation  ($2w[j]$). Consequently, the $3$ least significant digit slices affected by the truncation error are not implemented in the forthcoming cycles. For the last $\delta$ iterations, the modules for input and carry-save adder are not present and one bit reduction in the digit slice is due to the left shift operation of the residual. 
The input digits used in the proposed design are shown in clear fonts in Table.~\ref{tbl: olm_16}, whereas the pipelined design without precision reduction uses both clear and shaded digits. The $[4:2]$ adder takes two inputs from \emph{Selector} units and residuals $WS$ and $WC$, generating redundant vectors $VS$ and $VC$, the sum of which is shown as $v[j]$ in Table.~\ref{tbl: olm_16}.

\begin{table*}[!ht]
\renewcommand{\arraystretch}{1.3}
\caption{Example of radix-2 online multiplication for $n=16$ with reduced working precision $p=13$.}
	\begin{center}
\resizebox{1.0\textwidth}{!}{
\begin{tabular}{c|c|c|l|l|l|c|l|c}
\hline \hline
\multirow{2}{*}{$j$} & \multirow{2}{*}{$x_{j+4}$} & \multirow{2}{*}{$y_{j+4}$} & \multicolumn{1}{c|}{\multirow{2}{*}{$x[j]$}} & \multicolumn{1}{c|}{\multirow{2}{*}{$y[j+1]$}} & \multicolumn{1}{c|}{\multirow{2}{*}{$v[j]$}} & \multicolumn{2}{c|}{$p_{j+1}$} & \multirow{2}{*}{Error bound} \\ \cline{7-8}
 &  &  & \multicolumn{1}{c|}{} & \multicolumn{1}{c|}{} & \multicolumn{1}{c|}{} & SD & \multicolumn{1}{c|}{Conventional} &  \\ \hline \hline

-3 & 1 & $\overline{1}$ & 0.0\add{000000000000000} & 1.1\add{000000000000000} & 11.1111 & - & \multicolumn{1}{c|}{-} & - \\ 

-2 & 1 & 1 & 0.1\add{000000000000000} & 1.11\add{00000000000000} & 11.11101 & - & \multicolumn{1}{c|}{-} & - \\ 

-1 & 0 & $\overline{1}$ & 0.11\add{00000000000000} & 1.101\add{0000000000000} & 11.101110 & - &\multicolumn{1}{c|}{-} & - \\ 

0 & $\overline{1}$ & 1 & 0.110\add{0000000000000} & 1.1011\add{000000000000} & 11.1001001 & 0 & 0.0 & 2$^{-1}$ \\ 

1 & 0 & 0 & 0.1011\add{000000000000} & 1.10110\add{00000000000} & 11.00100100 & $\overline{1}$ & -0.25 & 2$^{-2}$ \\ 

2 & $\overline{1}$ & 0 & 0.10110\add{00000000000} & 1.101100\add{0000000000} & 00.010100100 & 0 & -0.25 & 2$^{-3}$ \\

3 & $\overline{1}$ & $\overline{1}$ & 0.101011\add{0000000000} & 1.1010111\add{000000000} & 00.1001100011 & 1 & -0.1875 & 2$^{-4}$ \\ 

4 & 0 & 1 & 0.1010101\add{000000000} & 1.10101111\add{00000000} & 11.01000110110 & $\overline{1}$ & -0.21875 & 2$^{-5}$ \\ 

5 & 1 & 0 & 0.10101010\add{00000000} & 1.101011110\add{0000000} & 00.100000110110 & 0 & -0.21875 & 2$^{-6}$ \\ 

6 & 1 & 1 & 0.101010101\add{0000000} & 1.1010111101\add{000000} & 01.0001000111111 & 1 & -0.2109375 & 2$^{-7}$ \\ 

7 & $\overline{1}$ & $\overline{1}$ & 0.1010101011\add{000000} & 1.10101111001\add{00000} & 00.00011000101101 & 0 & -0.2109375 & 2$^{-8}$ \\ 

8 & 0 & 1 & 0.10101010101\add{00000} & 1.101011110011\add{0000} & 00.010001101011110 & 0 & -0.2109375 & 2$^{-9}$ \\ 

9 & $\overline{1}$ & 1 & 0.101010101010\add{0000} & 1.1010111100111\add{000} & 00.1010110011100101 & 1 & -0.2099609375 & 2$^{-10}$ \\ 

10 & 1 & $\overline{1}$ & 0.101010101001\add{1000} & 1.1010111100110\add{100} & 11.0011101001011100 & $\overline{1}$ & -0.21044921875 & 2$^{-11}$ \\ 

11 & 0 & 0 & 0.101010101\add{0011100} & 1.1010111100\add{110100} & 00.0111010010100 & 0 & -0.21044921875 & 2$^{-12}$ \\ 

12 & 0 & $\overline{1}$ & 0.101010\add{1010011100} & 1.1010111\add{100110011} & 00.1101010000 & 1 & -0.2103271484375 & 2$^{-13}$ \\ 

13 & - & - & \multicolumn{1}{c|}{-} & \multicolumn{1}{c|}{-} & 11.101001100 & 0 & -0.2103271484375 & 2$^{-14}$ \\ 

14 & - & - & \multicolumn{1}{c|}{-} & \multicolumn{1}{c|}{-} & 11.01010000 & $\overline{1}$ & -0.210357666015625 & 2$^{-15}$ \\ 

15 & - & - & \multicolumn{1}{c|}{-} & \multicolumn{1}{c|}{-} & 00.1010000 & 1 & -0.2103424072265625 & 2$^{-16}$ \\ \hline \hline

\end{tabular}}
\end{center}
\label{tbl: olm_16}
\end{table*}

\subsection{Synthesis Results and Experimental Setup}
Conventional multipliers are implemented either sequentially, or using combinational approach. The sequential multiplier computes the result recursively using a single adder and produces the product in $n$ steps. The combinational implementation however is composed of several adders arranged to form either an array or tree for partial product reduction. The product is computed in $1$ cycle, however, the cycle time of combinational designs depends on the operand's width and is larger than the sequential designs. We use a sequential multiplier from \cite{bewick1994fast} and a linear array based combinational multiplier from \cite{baugh1973two}, for comparison with the proposed online designs.

Table.~\ref{Comparison3} shows the number of clock cycles required to multiply two $K_i$ vectors, where $i \in \{1,2,\ldots 8\}$, each $n$ bits wide, forming a stream. The conventional and non-pipelined online multipliers require corresponding number of clock cycles to complete the computation of full precision result $K_i$ vector after which, the computation of $K_{i+1}$ vector can be started. The pipelined online multipliers, however, after processing MSD of $K_1$, takes MSD of $K_2$ and MSD-1 of $K_1$ in the next cycle as depicted in Fig.~\ref{fig: Pipeline_Function}, thus reducing the total number of clock cycles to compute all $K$ vectors. It takes $n+\delta+1$ cycles to fill the pipeline and produce the output of first vector, after which, an output is produced in each clock cycle. $\delta_{ss}$ and $\delta_{sp}$ depicts the online delays for online serial-serial and online serial-parallel multipliers respectively. For large number of vectors i.e., $K>>n$, the delay to fill pipeline is negligible and therefore, the proposed pipelined designs have significant advantage over other designs. With the precision independence, short and fixed critical path, the proposed online arithmetic-based designs can be operated at higher frequency compared to conventional arithmetic based multiplier designs.  

\begin{table}[!ht]
\renewcommand{\arraystretch}{1.4}
\caption{Number of clock cycles required to compute $K$ = $8$ products of two input vectors with $n$ bits.}
\begin{center}
\resizebox{0.85\textwidth}{!}{
\begin{tabular}{l|c|c|c|c|c}
\hline \hline
\multirow{2}{*}{\textbf{Multiplier Type}} & \multirow{2}{*}{\textbf{Clock Cycles}} & \multicolumn{4}{c}{\textbf{$n$}} \\ \cline{3-6}
 &  & \textbf{8} & \textbf{16} & \textbf{24} & \textbf{32} \\ \hline \hline
Sequential \cite{bewick1994fast} & $n*K$& 64 & 128 & 192 & 256 \\
Combinational (Array) \cite{baugh1973two} & $K$& 8 & 8 & 8 & 8 \\
Non-Pipelined Online Serial-Serial \cite{ercegovac2004digital} & $(n+\delta_{ss}+1)*K$& 96 & 160 & 224 & 288 \\
Non-Pipelined Online Serial-Parallel & $(n+\delta_{sp}+1)*K$& 88 & 152 & 216 & 280 \\
Pipelined Online Serial-Serial  & $(n+\delta_{ss}+1)+(K-1)$& 19 & 27 & 35 & 43 \\
Pipelined Online Serial-Parallel  & $(n+\delta_{sp}+1)+(K-1)$& 18 & 26 & 34 & 42 \\
\hline \hline
\multicolumn{6}{l}{\footnotesize $\delta_{ss} = 3$; $\delta_{sp} = 2$} \\
\end{tabular}}
\end{center}	
\label{Comparison3}
\end{table}
The behavioral description of all the designs for $8$, $16$ and $32$ bits precision including non-pipelined and pipelined online multipliers as well as the conventional multipliers were written in Verilog. Their functional verification was done using ModelSim. 
The designs for all the multipliers were synthesized using Synopsys design compiler with GSCL $45$nm Liberty cell library from the Free45PDK. It was aimed to find the shortest critical path of each design therefore, all  designs were time constrained to obtain the maximum achievable frequency. 
Tables.~\ref{Comparison1}, \ref{Comparison1_16}, and \ref{Comparison1_32} present the post-synthesis results for online pipelined/non-pipelined serial-serial and serial-parallel multipliers along with conventional multiplier designs for $8, 16$, and $32$ bit precision respectively. As discussed earlier, this pipeline filling time is negligible for large number of vector computation and therefore, not considered while computing the performance and performance density. The evaluation of the designs has been presented for several parameters discussed in the following sections. 

\subsubsection{Period}
Several multiplier types require different number of clock cycles to produce the result according to the precision of operands and implementation. The multiplier implemented using combinational logic takes one clock cycle to produce the output, however, the cycle time or period of the clock varies correspondingly. Therefore, we present the results of the period to show the critical path. All the designs were time constrained to yield the smallest critical path, or in other words, the designs were executed at the maximum achievable frequency. It can be observed that the period for the conventional multipliers is dependent on the word size, whereas the online multipliers have smaller cycle time. The cycle time of the online multipliers remains constant when they are pipelined and is independent of word length, suggesting the opportunity to execute them at much higher frequency. The cycle time of online multiplier with both operands in serial is approximately $12\%$, $20\%$, and $92\%$ smaller for $8$, $16$, and $32$ bit sequential multiplier respectively. Comparing it with combinational multiplier, a reduction of approximately $58\%$, $113\%$, and $326\%$ is observed for $8$, $16$, and $32$ bit multipliers. Online serial-parallel multiplier has approximately $68\%$, $80\%$ and $188\%$ smaller cycle time for $8$, $16$, and $32$ bit designs of sequential multiplier. Comparing it with the combinational multiplier, a reduction of approximately $138\%$, $220\%$, and $540\%$ cycle time is observed for $8$, $16$, and $32$ bit multiplier designs respectively.

\subsubsection{Latency}
Latency of the online multipliers depend on the precision of the operands and the online delay. However, the inter-operation latency depends on $\delta$ only, which is fixed and small. For a series of online operations, the overall latency is the sum of online delays of the corresponding operation and is independent of precision. This implies that the use of online arithmetic based algorithms is even better for wider word sizes and long chains of data dependent operations.

\subsection{Power, Performance and Area (PPA)}

The proposed pipelined designs have been optimized to have savings in area and power. They account for higher area utilization and power consumption than the non-pipelined online and conventional multipliers. However, it is noteworthy that for $K$ number of vector multiplication, each pipeline stage is computing inputs from a distinct vector as shown in Fig.~\ref{fig: Pipeline_Function}, ultimately increasing the performance and performance density than the non-pipelined and conventional multipliers. The pipelined online multiplier with reduced working precision has $38$\% and $44$\% less power consumption and area utilization respectively than the pipelined online multiplier with full working precision design \cite{usman2021}. Both dynamic and static power of each designs have been considered and results of total power have been reported for $8$, $16$, and $32$ bit multiplier designs in Tables. \ref{Comparison1},  \ref{Comparison1_16}, and \ref{Comparison1_32} respectively.

The throughput of the multipliers in OPS (operations per second) has been reported as performance metric. The pipelined designs produce one vector per cycle in steady state, whereas the number of clock cycles to produce a vector output by the non-pipelined online designs depends on the word size hence increases linearly with the bit precision. An improvement of approximately $88\%$, $94\%$, and $98\%$ is observed for $8, 16$, and $32$ bit precision in pipelined online serial-serial multiplier compared with sequential multiplier respectively. Similarly, an improvement of approximately $36\%$, $53\%$, and $76\%$ is observed for $8, 16$, and $32$ bit precision, respectively, when compared with the performance of combinational multiplier. Pipelined serial-parallel multiplier shows even better performance, in particular it shows performance improvement of approximately $92\%$, $96\%$, and $98\%$ compared to the  $8, 16$, and $32$ bit precision sequential multipliers respectively. Furthermore, an improvement of approximately $58\%$, $68\%$, and $84\%$ is observed for $8, 16$, and $32$ bit precision respectively when compared with the performance of combinational multiplier.

\subsubsection{Energy-Delay Product (EDP)}

Energy delay product is useful metric that shows the delay of an operation times the energy consumed to perform that operation. Smaller values of energy-delay product suggest a more energy-efficient design \cite{horowitz1994low}. The results of EDP metric have been shown in Tables. \ref{Comparison1},  \ref{Comparison1_16}, and \ref{Comparison1_32} in the order of zepto-Joules ($10^{-21}$J).

\subsubsection{Performance Density}
Performance density is a useful metric to perceive the actual performance of the proposed designs. It is defined as the number of operations performed per unit area. The proposed  implementation results in an higher performance density compared to the conventional designs. In particular for $32$ bit precision, the pipelined online multiplier with both serial inputs has approximately $69\%$ and $95\%$ higher performance density than sequential and combinational multipliers respectively. Similarly, the pipelined online multiplier with one input in parallel shows $74\%$ and $96\%$ higher performance density than sequential and combinational multipliers for $32$ bit precision.

\begin{landscape}
\begin{table*}[!ht]
\renewcommand{\arraystretch}{2.4}
\caption{Synthesis results for several $8$ bit multipliers using Synopsys Design Compiler with GSCL $45$nm technology.}
\begin{center}
\resizebox{1.5\textwidth}{!}{
\begin{tabular}{l|c|c|c|c|c|c} \hline \hline
\textbf{Design} & \textbf{Sequential \cite{bewick1994fast}} & \textbf{Array \cite{baugh1973two}} & \textbf{\begin{tabular}[c]{@{}c@{}}Non-Pipelined \\ Online Serial-Serial \cite{ercegovac2004digital} \end{tabular}} & \textbf{\begin{tabular}[c]{@{}c@{}}Non-Pipelined \\ Online Serial-Parallel\end{tabular}} & \textbf{\begin{tabular}[c]{@{}c@{}}Pipelined Online \\ Serial-Serial\end{tabular}} & \textbf{\begin{tabular}[c]{@{}c@{}}Pipelined Online \\ Serial-Parallel\end{tabular}} \\ \hline \hline
Period (ns) 
& 0.84 
& 1.19 
& 0.75 
& 0.50
& 0.75 
& 0.50 \\ \hline

Latency (ns) 
& 8 cycles = 6.72 
& 1 cycle = 1.19 
& 11 cycles = 8.25 
& 10 cycles = 5.00 
& 11 cycles = 8.25 
& 10 cycles = 5.00 \\ \hline

Area ($\mu m^2$) 
& 1,174.94 
& 1,315.44 
& 1,614.39 
& 459.91 
& 5,174.5 
& 3,516.94 \\ \hline

Power ($mW$) 
& 0.91 
& 0.06 
& 1.71 
& 0.57 
& 5.38 
& 4.27 \\ \hline

EDP ($zJ$) 
& 0.64 
& 0.09 
& 0.96 
& 0.14 
& 0.37 
& 0.13 \\ \hline


Performance 
& \begin{tabular}[c]{@{}c@{}}1 vector/8 cycles \\ = 0.14 x $10^9$ OPS\end{tabular} 
& \begin{tabular}[c]{@{}c@{}}1 vector/1 cycle \\ =  0.84 x $10^9$ OPS\end{tabular} 
& \begin{tabular}[c]{@{}c@{}}1 vector/12 cycles \\ = 0.12 x $10^9$ OPS\end{tabular} 
& \begin{tabular}[c]{@{}c@{}}1 vector/11 cycles \\ = 0.20 x $10^9$ OPS\end{tabular} 
& \begin{tabular}[c]{@{}c@{}}1 vector/1 cycle \\ = 1.33 x $10^9$ OPS\end{tabular} 
& \begin{tabular}[c]{@{}c@{}}1 vector/1 cycle \\ =  2.00 x $10^9$ OPS\end{tabular} \\ \hline

Performance Density 
& 0.85E-03 OPs/1$\mu m^2$ 
& 0.76E-03 OPs/1$\mu m^2$ 
& 0.61E-03 OPs/1$\mu m^2$ 
& 2.17E-03 OPs/1$\mu m^2$ 
& 1.54E-03 OPs/1$\mu m^2$ 
& 2.27E-03 OPs/1$\mu m^2$ \\ \hline \hline

\multicolumn{7}{l}{\footnotesize *Assuming $\delta = 3$ for online serial-serial and $\delta = 2$ for online serial-parallel multipliers. } \\
 \multicolumn{7}{l}{\footnotesize OPS = Operations per second}\\
\multicolumn{7}{l}{\footnotesize OPs = Number of operations}
\end{tabular}}
\label{Comparison1}
\end{center}
\end{table*}
\end{landscape}

\begin{landscape}
\begin{table}[!ht]
\renewcommand{\arraystretch}{2.4}
\caption{Synthesis results for several $16$ bit multipliers using Synopsys Design Compiler with GSCL $45$nm technology.}
\begin{center}
\resizebox{1.5\textwidth}{!}{
\begin{tabular}{l|c|c|c|c|c|c} \hline \hline
\textbf{Design} & \textbf{Sequential \cite{bewick1994fast}} & \textbf{Array \cite{baugh1973two}} & \textbf{\begin{tabular}[c]{@{}c@{}}Non-Pipelined \\ Online Serial-Serial \cite{ercegovac2004digital}\end{tabular}} & \textbf{\begin{tabular}[c]{@{}c@{}}Non-Pipelined \\ Online Serial-Parallel\end{tabular}} & \textbf{\begin{tabular}[c]{@{}c@{}}Pipelined Online \\ Serial-Serial\end{tabular}} & \textbf{\begin{tabular}[c]{@{}c@{}}Pipelined Online \\ Serial-Parallel\end{tabular}} \\ \hline \hline

Period (ns) 
& 0.90
& 1.60
& 0.75
& 0.50
& 0.75
& 0.50 \\ \hline

Latency (ns)
& 16 cycles = 14.40 
& 1 cycle = 1.60
& 19 cycles = 14.25
& 18 cycles = 9.00
& 19 cycles = 15.25
& 18 cycles = 9.00 \\ \hline

Area ($\mu m^2$) 
& 2,604.15 
& 7,816.83 
& 2,458.66 
& 814.70 
& 16,408.14 
& 11,561.00 \\ \hline

Power ($mW$) 
& 1.80 
& 0.57 
& 2.40 
& 1.11 
& 16.88 
& 15.04 \\ \hline

EDP ($zJ$) 
& 1.46
& 1.46 
& 1.35 
& 0.27 
& 0.59 
& 0.23 \\ \hline


Performance 
& \begin{tabular}[c]{@{}l@{}}1 vector/17 cycles \\ = 0.06 x $10^9$ OPS\end{tabular} 
& \begin{tabular}[c]{@{}l@{}}1 vector/1 cycles \\ = 0.62 x $10^9$ OPS\end{tabular}
& \begin{tabular}[c]{@{}l@{}}1 vector/20 cycles \\ = 0.07 x $10^9$ OPS\end{tabular} 
& \begin{tabular}[c]{@{}l@{}}1 vector/19 cycles \\ = 0.11 x $10^9$ OPS\end{tabular} 
& \begin{tabular}[c]{@{}l@{}}1 vector/1 cycle \\ = 1.33 x $10^9$ OPS\end{tabular} 
& \begin{tabular}[c]{@{}l@{}}1 vector/1 cycle \\ =  2.00 x $10^9$ OPS\end{tabular} \\ \hline

Performance Density 
& 0.38E-03 OPs/1$\mu m^2$ 
& 0.12E-03 OPs/1$\mu m^2$ 
& 0.40E-03 OPs/1$\mu m^2$ 
& 1.23E-03 OPs/1$\mu m^2$ 
& 0.97E-03 OPs/1$\mu m^2$ 
& 1.38E-03 OPs/1$\mu m^2$ \\ \hline \hline

\multicolumn{7}{l}{\footnotesize *Assuming $\delta = 3$ for online serial-serial and $\delta = 2$ for online serial-parallel multipliers. } \\
 \multicolumn{7}{l}{\footnotesize OPS = Operations per second}\\
\multicolumn{7}{l}{\footnotesize OPs = Number of operations}
\end{tabular}}
\label{Comparison1_16}
\end{center}
\end{table}
\end{landscape}

\begin{landscape}
\begin{table*}[!ht]
\renewcommand{\arraystretch}{2.4}
\caption{Synthesis results for several $32$ bit multipliers using Synopsys Design Compiler with GSCL $45$nm technology.}
\begin{center}
\resizebox{1.5\textwidth}{!}{
\begin{tabular}{l|c|c|c|c|c|c} \hline \hline
\textbf{Design} & \textbf{Sequential \cite{bewick1994fast}} & \textbf{Array \cite{baugh1973two} } & \textbf{\begin{tabular}[c]{@{}c@{}}Non-Pipelined \\ Online Serial-Serial \cite{ercegovac2004digital}\end{tabular}} & \textbf{\begin{tabular}[c]{@{}c@{}}Non-Pipelined \\ Online Serial-Parallel\end{tabular}} & \textbf{\begin{tabular}[c]{@{}c@{}}Pipelined Online \\ Serial-Serial\end{tabular}} & \textbf{\begin{tabular}[c]{@{}c@{}}Pipelined Online \\ Serial-Parallel\end{tabular}} \\ \hline \hline

Period (ns) 
& 1.44
& 3.20
& 0.75
& 0.50
& 0.75
& 0.50 \\ \hline

Latency (ns)
& 32 cycles = 46.08
& 1 cycle = 3.20
& 35 cycles = 26.25
& 34 cycles = 17.00
& 35 cycles = 26.25
& 34 cycles = 17.00 \\ \hline

Area ($\mu m^2$) 
& 4,807.50 
& 33,626.65
& 4,567.22 
& 1,530.40 
& 49,365.89 
& 39,606.71 \\ \hline

Power ($mW$) 
& 2.12 
& 3.10 
& 4.41 
& 2.13 
& 59.91 
& 55.75 \\ \hline

EDP ($zJ$) 
& 4.40
& 31.8 
& 2.48 
& 0.53 
& 1.50 
& 0.43 \\ \hline


Performance 
&\begin{tabular}[c]{@{}c@{}}1 vector/33 cycles \\ =  0.02 x $10^9$ OPS\end{tabular} 
& \begin{tabular}[c]{@{}c@{}}1 vector/1 cycle \\ = 0.31 x $10^9$ OPS\end{tabular} 
& \begin{tabular}[c]{@{}c@{}}1 vector/36 cycles \\ = 0.03 x $10^9$ OPS\end{tabular} 
& \begin{tabular}[c]{@{}c@{}}1 vector/35 cycles \\ = 0.05 x $10^9$ OPS\end{tabular} 
& \begin{tabular}[c]{@{}c@{}}1 vector/1 cycle \\ = 1.33 x $10^9$ OPS\end{tabular} 
& \begin{tabular}[c]{@{}c@{}}1 vector/1 cycle \\ =  2.00 x $10^9$ OPS\end{tabular} \\ \hline

Performance Density 
& 2.08E-04 OPs/1$\mu m^2$ 
& 0.29E-04 OPs/1$\mu m^2$ 
& 2.19E-04 OPs/1$\mu m^2$ 
& 6.53E-04 OPs/1$\mu m^2$ 
& 6.48E-04 OPs/1$\mu m^2$ 
& 8.08E-04 OPs/1$\mu m^2$ \\ \hline \hline

\multicolumn{7}{l}{\footnotesize *Assuming $\delta = 3$ for online serial-serial and $\delta = 2$ for online serial-parallel multipliers. } \\
 \multicolumn{7}{l}{\footnotesize OPS = Operations per second}\\
\multicolumn{7}{l}{\footnotesize OPs = Number of operations}
\end{tabular}}
\label{Comparison1_32}
\end{center}
\end{table*}
\end{landscape}




\section{Conclusion}\label{sec: conclusion}
In this paper, we present online arithmetic based serial-serial and serial-parallel multipliers which have been pipelined such that in a steady state, one vector is generated in each cycle. The properties of the online arithmetic not only allows massive pipelining of the successive operations regardless of the data dependency, the signal activity of the algorithms can also be reduced. The digit serial nature of the online arithmetic and possibility to reduce the maximum working precision, due to which $n$ bit precision result can be obtained by employing $p<n$ bits, manifest the reduction of active slices and signal activities which results in saving power and area during implementation. Several precision multipliers are proposed and compared with the conventional online multipliers. The proposed designs have been synthesized using Synopsys design compiler with GSCL $45$nm technology. Results show that the proposed designs produce accurate results with higher throughput and have better performance density compared to other designs. In future, we shall utilize the proposed designs to interface with the online arithmetic based adder to present kernels including matrix multiplication, FFT, sum-of-products, etc., for several real World applications. 




\bigskip

\bibliography{sn-bibliography}


\begin{thebibliography}{20}
\ifx \bisbn   \undefined \def \bisbn  #1{ISBN #1}\fi
\ifx \binits  \undefined \def \binits#1{#1}\fi
\ifx \bauthor  \undefined \def \bauthor#1{#1}\fi
\ifx \batitle  \undefined \def \batitle#1{#1}\fi
\ifx \bjtitle  \undefined \def \bjtitle#1{#1}\fi
\ifx \bvolume  \undefined \def \bvolume#1{\textbf{#1}}\fi
\ifx \byear  \undefined \def \byear#1{#1}\fi
\ifx \bissue  \undefined \def \bissue#1{#1}\fi
\ifx \bfpage  \undefined \def \bfpage#1{#1}\fi
\ifx \blpage  \undefined \def \blpage #1{#1}\fi
\ifx \burl  \undefined \def \burl#1{\textsf{#1}}\fi
\ifx \doiurl  \undefined \def \doiurl#1{\url{https://doi.org/#1}}\fi
\ifx \betal  \undefined \def \betal{\textit{et al.}}\fi
\ifx \binstitute  \undefined \def \binstitute#1{#1}\fi
\ifx \binstitutionaled  \undefined \def \binstitutionaled#1{#1}\fi
\ifx \bctitle  \undefined \def \bctitle#1{#1}\fi
\ifx \beditor  \undefined \def \beditor#1{#1}\fi
\ifx \bpublisher  \undefined \def \bpublisher#1{#1}\fi
\ifx \bbtitle  \undefined \def \bbtitle#1{#1}\fi
\ifx \bedition  \undefined \def \bedition#1{#1}\fi
\ifx \bseriesno  \undefined \def \bseriesno#1{#1}\fi
\ifx \blocation  \undefined \def \blocation#1{#1}\fi
\ifx \bsertitle  \undefined \def \bsertitle#1{#1}\fi
\ifx \bsnm \undefined \def \bsnm#1{#1}\fi
\ifx \bsuffix \undefined \def \bsuffix#1{#1}\fi
\ifx \bparticle \undefined \def \bparticle#1{#1}\fi
\ifx \barticle \undefined \def \barticle#1{#1}\fi
\bibcommenthead
\ifx \bconfdate \undefined \def \bconfdate #1{#1}\fi
\ifx \botherref \undefined \def \botherref #1{#1}\fi
\ifx \url \undefined \def \url#1{\textsf{#1}}\fi
\ifx \bchapter \undefined \def \bchapter#1{#1}\fi
\ifx \bbook \undefined \def \bbook#1{#1}\fi
\ifx \bcomment \undefined \def \bcomment#1{#1}\fi
\ifx \oauthor \undefined \def \oauthor#1{#1}\fi
\ifx \citeauthoryear \undefined \def \citeauthoryear#1{#1}\fi
\ifx \endbibitem  \undefined \def \endbibitem {}\fi
\ifx \bconflocation  \undefined \def \bconflocation#1{#1}\fi
\ifx \arxivurl  \undefined \def \arxivurl#1{\textsf{#1}}\fi
\csname PreBibitemsHook\endcsname

\bibitem{liu2017design}
\begin{barticle}
\bauthor{\bsnm{Liu}, \binits{W.}},
\bauthor{\bsnm{Qian}, \binits{L.}},
\bauthor{\bsnm{Wang}, \binits{C.}},
\bauthor{\bsnm{Jiang}, \binits{H.}},
\bauthor{\bsnm{Han}, \binits{J.}},
\bauthor{\bsnm{Lombardi}, \binits{F.}}:
\batitle{Design of approximate radix-4 booth multipliers for error-tolerant
  computing}.
\bjtitle{IEEE Transactions on Computers}
\bvolume{66}(\bissue{8}),
\bfpage{1435}--\blpage{1441}
(\byear{2017})
\end{barticle}
\endbibitem

\bibitem{lin2019novel}
\begin{barticle}
\bauthor{\bsnm{Lin}, \binits{J.-F.}},
\bauthor{\bsnm{Chan}, \binits{C.-Y.}},
\bauthor{\bsnm{Yu}, \binits{S.-W.}}:
\batitle{Novel low voltage and low power array multiplier design for iot
  applications}.
\bjtitle{Electronics}
\bvolume{8}(\bissue{12}),
\bfpage{1429}
(\byear{2019})
\end{barticle}
\endbibitem

\bibitem{ercegovac2017left}
\begin{bchapter}
\bauthor{\bsnm{Ercegovac}, \binits{M.D.}}:
\bctitle{On left-to-right arithmetic}.
In: \bbtitle{2017 51st Asilomar Conference on Signals, Systems, and Computers},
pp. \bfpage{750}--\blpage{754}
(\byear{2017}).
\bcomment{IEEE}
\end{bchapter}
\endbibitem

\bibitem{ercegovac2020}
\begin{bchapter}
\bauthor{\bsnm{Ercegovac}, \binits{M.D.}}:
\bctitle{On reducing module activities in online arithmetic operations}.
In: \bbtitle{2020 54th Asilomar Conference on Signals, Systems, and Computers},
pp. \bfpage{524}--\blpage{528}
(\byear{2020}).
\bcomment{IEEE}
\end{bchapter}
\endbibitem

\bibitem{villalba2011radix}
\begin{barticle}
\bauthor{\bsnm{Villalba}, \binits{J.}},
\bauthor{\bsnm{Lang}, \binits{T.}},
\bauthor{\bsnm{Hormigo}, \binits{J.}}:
\batitle{Radix-2 multioperand and multiformat streaming online addition}.
\bjtitle{IEEE Transactions on Computers}
\bvolume{61}(\bissue{6}),
\bfpage{790}--\blpage{803}
(\byear{2011})
\end{barticle}
\endbibitem

\bibitem{elshafei2009hardware}
\begin{botherref}
\oauthor{\bsnm{Elshafei}, \binits{A.-R.}}:
Hardware online multiplication-division: A design and performance study.
PhD thesis,
King Fahd University of Petroleum and Minerals
(2009)
\end{botherref}
\endbibitem

\bibitem{huang2001fpga}
\begin{bchapter}
\bauthor{\bsnm{Huang}, \binits{Z.}},
\bauthor{\bsnm{Ercegovac}, \binits{M.D.}}:
\bctitle{Fpga implementation of pipelined on-line scheme for 3-d vector
  normalization}.
In: \bbtitle{The 9th Annual IEEE Symposium on Field-Programmable Custom
  Computing Machines (FCCM'01)},
pp. \bfpage{61}--\blpage{70}
(\byear{2001}).
\bcomment{IEEE}
\end{bchapter}
\endbibitem

\bibitem{galli2001design}
\begin{botherref}
\oauthor{\bsnm{Galli}, \binits{R.}}:
Design and evaluation of on-line arithmetic modules and networks for signal
  processing applications on fpgas.
Master's thesis,
Oregon State University
(2001)
\end{botherref}
\endbibitem

\bibitem{sinky2004design}
\begin{bchapter}
\bauthor{\bsnm{Sinky}, \binits{M.H.}},
\bauthor{\bsnm{Tenca}, \binits{A.F.}},
\bauthor{\bsnm{Shantilal}, \binits{A.C.}},
\bauthor{\bsnm{Lucchese}, \binits{L.}}:
\bctitle{Design of a color image processing algorithm using online arithmetic
  modules}.
In: \bbtitle{Advanced Signal Processing Algorithms, Architectures, and
  Implementations XIV},
vol. \bseriesno{5559},
pp. \bfpage{79}--\blpage{90}
(\byear{2004}).
\bcomment{International Society for Optics and Photonics}
\end{bchapter}
\endbibitem

\bibitem{zhao2016efficient}
\begin{bchapter}
\bauthor{\bsnm{Zhao}, \binits{Y.}},
\bauthor{\bsnm{Wickerson}, \binits{J.}},
\bauthor{\bsnm{Constantinides}, \binits{G.A.}}:
\bctitle{An efficient implementation of online arithmetic}.
In: \bbtitle{2016 International Conference on Field-Programmable Technology
  (FPT)},
pp. \bfpage{69}--\blpage{76}
(\byear{2016}).
\bcomment{IEEE}
\end{bchapter}
\endbibitem

\bibitem{shi2014efficient}
\begin{bchapter}
\bauthor{\bsnm{Shi}, \binits{K.}},
\bauthor{\bsnm{Boland}, \binits{D.}},
\bauthor{\bsnm{Constantinides}, \binits{G.A.}}:
\bctitle{Efficient fpga implementation of digit parallel online arithmetic
  operators}.
In: \bbtitle{2014 International Conference on Field-Programmable Technology
  (FPT)},
pp. \bfpage{115}--\blpage{122}
(\byear{2014}).
\bcomment{IEEE}
\end{bchapter}
\endbibitem

\bibitem{usman2021}
\begin{bchapter}
\bauthor{\bsnm{Usman}, \binits{M.}},
\bauthor{\bsnm{Lee}, \binits{J.-A.}},
\bauthor{\bsnm{Ercegovac}, \binits{M.D.}}:
\bctitle{Multiplier with reduced activities and minimized interconnect for
  inner product arrays}.
In: \bbtitle{2021 55th Asilomar Conference on Signals, Systems, and Computers},
pp. \bfpage{1}--\blpage{5}
(\byear{2021}).
\bcomment{IEEE}
\end{bchapter}
\endbibitem

\bibitem{tangtrakul1996signed}
\begin{bchapter}
\bauthor{\bsnm{Tangtrakul}, \binits{A.}},
\bauthor{\bsnm{Yeung}, \binits{B.}},
\bauthor{\bsnm{Cook}, \binits{T.A.}}:
\bctitle{Signed-digit online floating-point arithmetic for fpgas}.
In: \bbtitle{High-Speed Computing, Digital Signal Processing, and Filtering
  Using Reconfigurable Logic},
vol. \bseriesno{2914},
pp. \bfpage{2}--\blpage{13}
(\byear{1996}).
\bcomment{International Society for Optics and Photonics}
\end{bchapter}
\endbibitem

\bibitem{dormiani2005design}
\begin{bchapter}
\bauthor{\bsnm{Dormiani}, \binits{P.}},
\bauthor{\bsnm{Omoto}, \binits{D.}},
\bauthor{\bsnm{Adharapurapu}, \binits{P.}},
\bauthor{\bsnm{Ercegovac}, \binits{M.D.}}:
\bctitle{A design of online scheme for evaluation of multinomials}.
In: \bbtitle{Advanced Signal Processing Algorithms, Architectures, and
  Implementations XV},
vol. \bseriesno{5910},
p. \bfpage{59100}
(\byear{2005}).
\bcomment{International Society for Optics and Photonics}
\end{bchapter}
\endbibitem

\bibitem{ercegovac1987fly}
\begin{botherref}
\oauthor{\bsnm{Ercegovac}, \binits{M.D.}},
\oauthor{\bsnm{Lang}, \binits{T.}}:
On-the-fly conversion of redundant into conventional representations.
IEEE Transactions on Computers
(7),
895--897
(1987)
\end{botherref}
\endbibitem

\bibitem{ercegovac2004digital}
\begin{bbook}
\bauthor{\bsnm{Ercegovac}, \binits{M.D.}},
\bauthor{\bsnm{Lang}, \binits{T.}}:
\bbtitle{Digital Arithmetic}.
\bpublisher{Morgan Kaufmann Publishers},
\blocation{San Francisco, CA, USA}
(\byear{2004})
\end{bbook}
\endbibitem

\bibitem{usmanthesis}
\begin{botherref}
\oauthor{\bsnm{Usman}, \binits{M.}}:
Energy-efficient online arithmetic in domain-specific accelerators for deep
  learning applications.
PhD thesis,
Chosun University
(2022)
\end{botherref}
\endbibitem

\bibitem{bewick1994fast}
\begin{botherref}
\oauthor{\bsnm{Bewick}, \binits{G.W.}}:
Fast multiplication: Algorithms and implementation.
PhD thesis,
Stanford University
(1994)
\end{botherref}
\endbibitem

\bibitem{baugh1973two}
\begin{barticle}
\bauthor{\bsnm{Baugh}, \binits{C.R.}},
\bauthor{\bsnm{Wooley}, \binits{B.A.}}:
\batitle{A two's complement parallel array multiplication algorithm}.
\bjtitle{IEEE Transactions on computers}
\bvolume{100}(\bissue{12}),
\bfpage{1045}--\blpage{1047}
(\byear{1973})
\end{barticle}
\endbibitem

\bibitem{horowitz1994low}
\begin{bchapter}
\bauthor{\bsnm{Horowitz}, \binits{M.}},
\bauthor{\bsnm{Indermaur}, \binits{T.}},
\bauthor{\bsnm{Gonzalez}, \binits{R.}}:
\bctitle{Low-power digital design}.
In: \bbtitle{Proceedings of 1994 IEEE Symposium on Low Power Electronics},
pp. \bfpage{8}--\blpage{11}
(\byear{1994}).
\bcomment{IEEE}
\end{bchapter}
\endbibitem

\end{thebibliography}

\end{document}